\documentclass[sigconf]{acmart}

\AtBeginDocument{%
  }

\copyrightyear{2025}
\acmYear{2025}
\setcopyright{acmlicensed}\acmConference[KDD '25]{Proceedings of the 31st ACM SIGKDD Conference on Knowledge Discovery and Data Mining V.2}{August 3--7, 2025}{Toronto, ON, Canada}
\acmBooktitle{Proceedings of the 31st ACM SIGKDD Conference on Knowledge Discovery and Data Mining V.2 (KDD '25), August 3--7, 2025, Toronto, ON, Canada}
\acmDOI{10.1145/3711896.3736963}
\acmISBN{979-8-4007-1454-2/2025/08}

\usepackage{hyperref}
\usepackage{url}
\usepackage{amsfonts}
\usepackage{graphicx}
\usepackage{tcolorbox}
\usepackage{xcolor}
\usepackage{tikz}
\usepackage{booktabs}
\usepackage{mdframed}
\usepackage{pifont}
\usepackage{xspace}
\usepackage{multirow}
\usepackage{wrapfig}
\usepackage{bm}
\usepackage{nicefrac}
\usepackage{microtype}
\usepackage{subfigure}
\usepackage{amsmath}

\usepackage{amssymb}

\usepackage{makeidx}    
\usepackage{multicol}
\usepackage{balance}
\usepackage{enumitem}
\usepackage{array}
\usepackage{comment}
\usepackage{caption}
\usepackage{adjustbox}
\usepackage{amsthm}
\usepackage{mwe}

\makeindex              

\newcommand{\datasetFont}{\text}
\newcommand{\ours}{\datasetFont{\texttt{FLanS}}\xspace}

\definecolor{darkpink}{rgb}{0.8, 0.2, 0.5}
\definecolor{lightblue}{rgb}{0.4, 0.529, 0.639}
\usepackage{setspace}
\newcolumntype{P}[1]{>{\centering\arraybackslash}p{#1}}
\definecolor{grey}{gray}{0.5}
\definecolor{custompurple}{HTML}{5b457a}

\begin{document}

\title{FLanS: A Foundation Model for Free-Form Language-based Segmentation in Medical Images}


\author{Longchao Da}
\authornote{Authors contributed equally to this research. Work was done during Longchao's internship at GE-Healthcare.}
\orcid{0009-0000-8631-9634}
\affiliation{%
  \institution{Arizona State University}
  \city{Tempe}
  \state{AZ}
  \country{USA}
}
\email{longchao@asu.edu}

\author{Rui Wang}\authornotemark[1]
\authornote{Work done prior to joining Amazon.}
\affiliation{%
  \institution{Amazon Web Serives}
  \city{Seattle}
  \state{WA}
  \country{USA}}
\email{rwngamz@amazon.com}

\author{Xiaojian Xu}
\affiliation{%
  \institution{GE-Healthcare}
  \city{Bellevue}
  \state{WA}
  \country{USA}
}
\author{Parminder Bhatia}
\affiliation{%
  \institution{GE-Healthcare}
  \city{Bellevue}
  \state{WA}
  \country{USA}}
  
\author{Taha Kass-Hout}
\affiliation{%
 \institution{GE-Healthcare}
 \city{Bellevue}
 \state{WA}
 \country{USA}}

\author{Hua Wei}
\affiliation{%
  \institution{Arizona State University}
  \city{Tempe}
  \state{AZ}
  \country{USA}}

\author{Cao Xiao}
\affiliation{%
  \institution{GE-Healthcare}
  \city{Bellevue}
  \country{USA}}
\email{cao.xiao@gehealthcare.com}

\renewcommand{\shortauthors}{Da et al.}

\begin{abstract}
  Medical imaging is crucial for diagnosing a patient’s health condition, and accurate segmentation of these images is essential for isolating regions of interest to ensure precise diagnosis and treatment planning. Existing methods primarily rely on bounding boxes or point-based prompts, while few have explored text-related prompts, despite clinicians often describing their observations and instructions in natural language. To address this gap, we first propose a RAG-based free-form text prompt generator, that leverages the domain corpus to generate diverse and realistic descriptions. Then, we introduce \texttt{FLanS}, a novel medical image segmentation model that handles various free-form text prompts, including professional anatomy-informed queries, anatomy-agnostic position-driven queries, and anatomy-agnostic size-driven queries. Additionally, our model also incorporates a symmetry-aware canonicalization module to ensure consistent, accurate segmentations across varying scan orientations and reduce confusion between the anatomical position of an organ and its appearance in the scan. \texttt{FLanS} is trained on a large-scale dataset of over 100k medical images from 7 public datasets. Comprehensive experiments demonstrate the model’s superior language understanding and segmentation precision, along with a deep comprehension of the relationship between them, outperforming SOTA baselines on both in-domain and out-of-domain datasets. For code and models, please find them in the released project page by clicking \href{https://longchaoda.github.io/segmentAsYouWish.github.io/}{\textcolor{blue}{here}}.
\end{abstract}

\ccsdesc[500]{Computing methodologies~Medical image segmentation}
\ccsdesc[300]{Computing methodologies~Neural networks}
\ccsdesc[100]{Computing methodologies~Language models}
\ccsdesc[100]{Information systems~Retrieval augmented generation}

\keywords{Foundation Model, Medical Image, Language Based Segmentation}


\maketitle
\section{Introduction}
\label{sec:intro}

Medical imaging is crucial in healthcare, providing clinicians with the ability to visualize and assess anatomical structures for both diagnosis and treatment. Organ segmentation is vital for numerous clinical applications, including surgical planning and disease progression monitoring \citep{wang2022medical, du2020medical, shamshad2023transformers}. However, accurately segmenting organs and tissues from these medical images, i.e., \underline{m}edical \underline{i}mage \underline{s}egmentation (MIS), remains a significant challenge due to the variability in patient positioning, imaging techniques, and anatomical structures \citep{pham2000current, xiao2021introduction}. Recent advancements in large foundation models, such as Segment Anything Model (SAM) \citep{kirillov2023segment} and MedSAM \citep{wu2023medical}, have shown promise in achieving more accurate and faster MIS. These models often require the users to input a predefined category name, a box,  or a point as a prompt. However, in real-world scenarios, clinicians often rely on natural language commands to interact with medical images, such as ``\textit{Highlight the right kidney}" or ``\textit{Segment the largest organ}". An accurate segmentation model with flexible text comprehension capability is therefore essential for a wide range of clinical applications.  

\begin{figure}[h!]
    \centering
    \includegraphics[width=\linewidth]{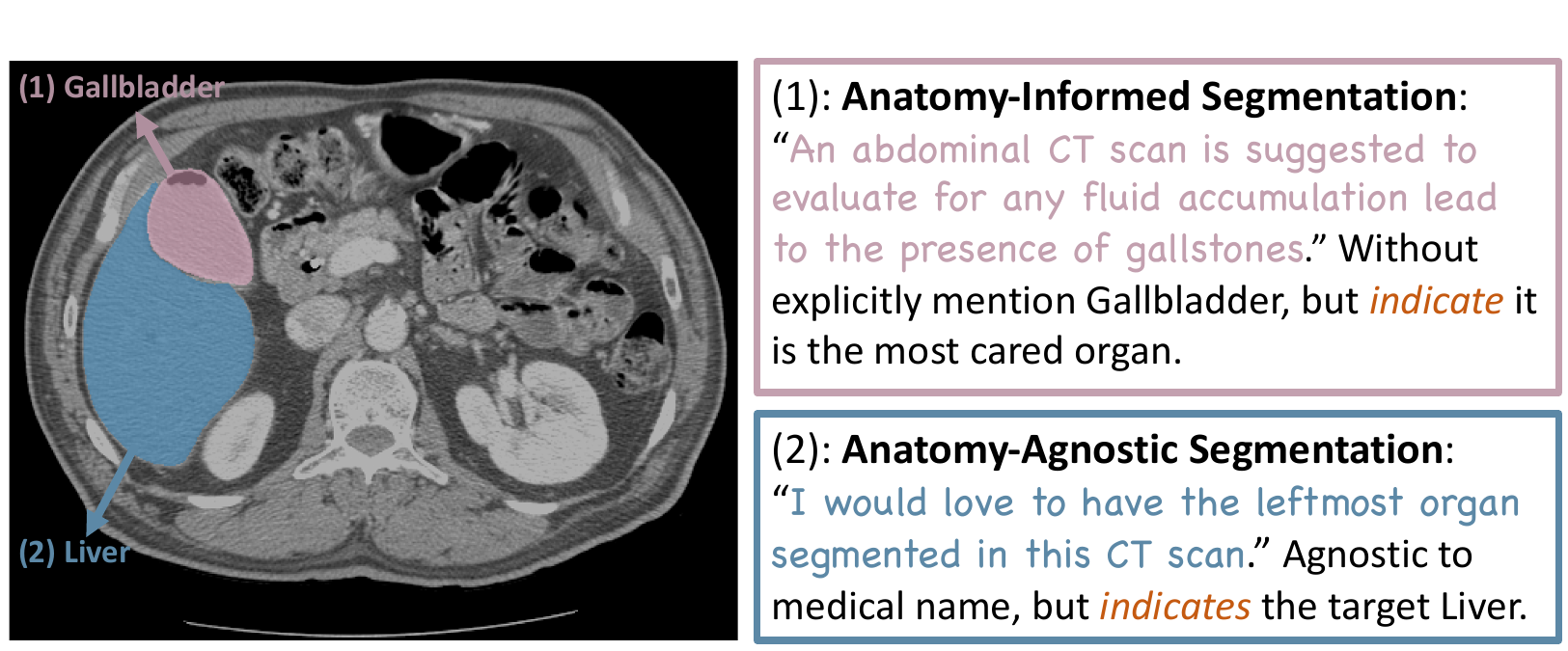}
    \caption{Examples of two types of language-prompted segmentation. (1) \textit{\textcolor{darkpink}{A professional diagnosis snippet from EMR uses medically relevant information to indicate that the most important segmentation area is the \textbf{Gallbladder}.}} (2) \textit{\textcolor{lightblue}{This description is agnostic to the medical name of the organ but identifies the segmentation target based on positional semantics: the \textbf{Liver}.}}}
    \label{fig:example1}
\end{figure}

\textbf{The first challenge} lies in the development of a segmentation model that can handle text prompts, offering greater flexibility and adaptability in real-world clinical environments. Unlike traditional models that rely on bounding boxes (Bboxes) or point prompts, this method should allow clinicians to use \textit{free-form} natural language commands and streamline the diagnostic process by enabling intuitive, verbal interactions. For \textit{free-form} text, we provide two conceptual definitions as follows:
 (1) \textit{Anatomy-Informed Segmentation}, where the user has explicit knowledge of the organ or relevant pathology to be segmented; (2) \textit{Anatomy-Agnostic Segmentation},  where the user lacks medical knowledge about a specific organ or CT scan and hence queries based on positional information, organ sizes or other visible characteristics. \textbf{This scenario is more common for education purposes that individuals (like students) without formal medical training. } An exemplar illustration is shown in Fig.~\ref{fig:example1}.~\footnote{All of the images in this paper are best viewed in color.}







To learn a free-form text-supportive MIS model, text prompt generation towards the groundtruth mask is a primary step. Instead of using labor-intensive manual labeling to match with the masks, we propose a retrieval augmented generation (RAG)~\citep{lewis2020retrieval} fashion method that automates text query generation using corpus embeddings collected from three resources (clinical expert records, non-expert queries, and synthetic queries). This approach guarantees that the generated query prompts capture various forms of language use across different demographic groups. Based on the text queries, we propose \textbf{\ours{}}, a \textbf{f}ree-form \textbf{lan}guage-based \textbf{s}egmentation model that can accurately interpret and respond to \textit{free-form} prompts either professional or straightforward, ensuring accurate segmentation across a variety of query scenarios.

\begin{figure}[t!]
    \centering
    \captionof{table}{\ours{} uniquely supports all prompt types, including free-form text, and is symmetry-aware.}
    \label{tab:salesman}
    \scriptsize
    \resizebox{0.48\textwidth}{!}{
    \begin{tabular}
    {P{2.85cm}|P{0.3cm}P{0.3cm}P{0.3cm}P{0.3cm}P{0.9cm}}
    \toprule
    \textbf{Model}  & \multicolumn{4}{c}{\textbf{Prompt Type}} & \textbf{Symmetry Aware} \\ \cline{2-5}
                    & \textit{Label} & \textit{Point} & \textit{Bbox} & \textit{Text} & \\ \hline
    SAM-U~\citep{deng2023sam}                     & \textcolor{red}{\ding{56}}  & \textcolor{green}{\ding{52}} & \textcolor{green}{\ding{52}} & \textcolor{red}{\ding{56}} & \textcolor{red}{\ding{56}}  \\ 
    SAMed~\citep{zhang2023customized}             & \textcolor{red}{\ding{56}}  & \textcolor{green}{\ding{52}}  & \textcolor{green}{\ding{52}}   & \textcolor{red}{\ding{56}} & \textcolor{red}{\ding{56}}  \\
    AutoSAM~\citep{hu2023efficiently}             & \textcolor{red}{\ding{56}}  & \textcolor{green}{\ding{52}} & \textcolor{green}{\ding{52}}  & \textcolor{red}{\ding{56}} & \textcolor{red}{\ding{56}}  \\
    MedSAM~\citep{ma2024segment}                  & \textcolor{red}{\ding{56}}  & \textcolor{green}{\ding{52}} & \textcolor{green}{\ding{52}} & \textcolor{red}{\ding{56}} & \textcolor{red}{\ding{56}}  \\
    MSA~\citep{wu2023medical}                     & \textcolor{red}{\ding{56}}  & \textcolor{green}{\ding{52}}& \textcolor{green}{\ding{52}}  & \textcolor{red}{\ding{56}} & \textcolor{red}{\ding{56}}  \\ 
    Universal ~\citep{liu2023clip} 
     &\textcolor{green}{\ding{52}} &
      \textcolor{red}{\ding{56}} &
      \textcolor{red}{\ding{56}} &
      \textcolor{red}{\ding{56}} &
      \textcolor{red}{\ding{56}} \\
    \hline
    \textbf{\ours (ours) }                                 & \textcolor{green}{\ding{52}}& \textcolor{green}{\ding{52}}& \textcolor{green}{\ding{52}} & \textcolor{green}{\ding{52}}& \textcolor{green}{\ding{52}} \\ 
    \bottomrule
    \end{tabular}
    }
    \includegraphics[width=0.49\textwidth]{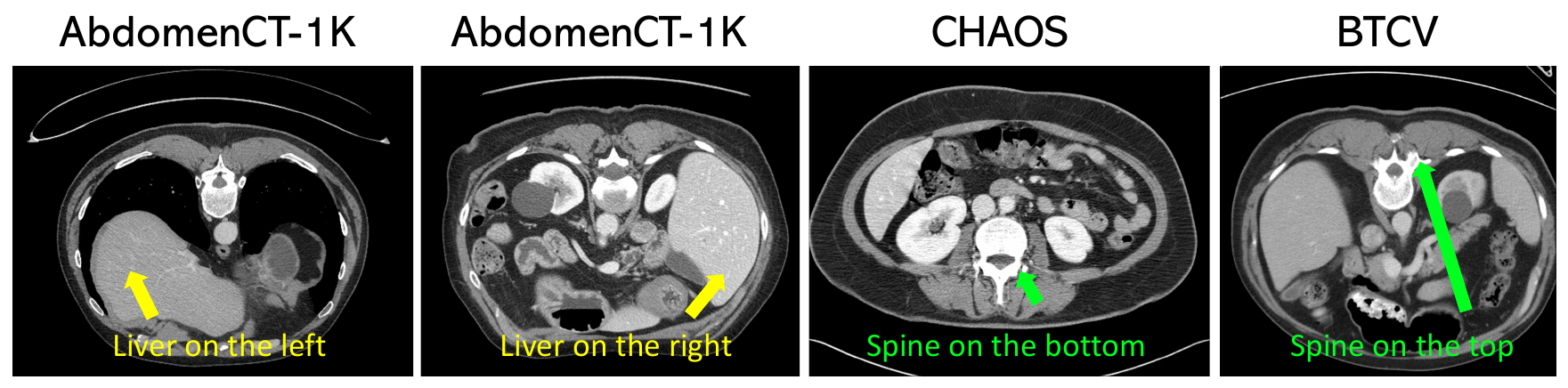}
    \caption{Example CT images from different datasets show variations in orientation, which highlight the need for a symmetry-aware (equivariant) model to ensure consistent segmentation performance across diverse scan orientations.}
    \label{fig: orientations}
\end{figure}

\textbf{Another challenge} in text-based medical imaging segmentation arises from the variability in scan orientation. Factors such as patient positioning (e.g., supine vs. prone), different imaging planes (axial, coronal, sagittal), reconstruction algorithms and settings, and the use of portable imaging devices in emergency settings can cause organs to appear in unexpected locations or orientations. The scan orientations even differ between well-preprocessed datasets, such as AbdomenCT-1K \citep{Ma-2021-AbdomenCT-1K} and BTCV \citep{gibson_2018_1169361}, as shown in Fig.~\ref{fig: orientations}. This variability can confuse segmentation models, making it difficult to distinguish between the anatomical position of an organ and its appearance in a scan. For instance, the right kidney may appear either on the left or the right side of a rotated scan, leading to inaccurate segmentations. To address this challenge, we integrate the symmetry-aware canonicalization module as a crucial step in our model architecture \citep{kaba2022equivariance, mondal2023equivariant}, which ensures the model produces consistent segmentations regardless of the scan's orientation, enhancing its accuracy across diverse medical images \citep{Cohen2016Group, weiler2019e2cnn}. 
Additionally, incorporating symmetry improves sample efficiency and generalizability, which is well-suited for medical imaging tasks where labeled datasets are limited \citep{wang2022surprising, wang2021incorporating, zhu2022grasp, thomas2018tensor}.

Our key contributions in this paper are summarized as follows:
\begin{itemize}[leftmargin=15pt, itemsep=0.5pt]
    \item We employ RAG techniques to generate free-form text prompts for various anatomical structures, incorporating both anatomy-informed and anatomy-agnostic queries. Derived from the vectorized embeddings of clinical reports, the generated query data adopts realistic tones and natural word usage.

    \item We introduce \textbf{\ours{}}, a novel medical image segmentation model that demonstrates a deep understanding of the relationship between text descriptions and medical images. It uniquely supports free-form text segmentation and employs a symmetry-aware canonicalization module to handle variability in scan orientation, as illustrated in Table.~\ref{tab:salesman}.


    \item We demonstrate the effectiveness of \textbf{\ours{}} across diverse anatomical structures and clinical scenarios using both in-domain and out-of-domain datasets. Additionally, we conduct comprehensive ablation studies to validate the contribution of each component in our model design.
\end{itemize}

\section{Related Work}

\paragraph{Medical Image Segmentation}
Medical image segmentation (MIS) aims at accurately delineating anatomical structures in medical images. Traditionally, MIS methods tend to segment the correct regions from an image that accurately reflects the input query~\citep{azad2024medical}. The researchers improve the performance of MIS methods by either optimizing segmentation network design for improving feature representations~\citep{encodedecode, zhao2017pyramid, chen2018drinet, gu2020net}, or improving optimization strategies, e.g., proposing better loss functions to address class imbalance or refining uncertain pixels from high-frequency regions to improve the segmentation quality~\citep{xue2020shape, shi2021marginal, you2022simcvd}.  However, they require a pre-known medical region from the user as an input for segmentation on where it is expected to be segmented and a precise match between the segment's name and the labels used in the training set, restricting their flexibility in real-world application. Another category of methods are SAM-based approaches~\citep{kirillov2023segment, ma2024segment, zhu2024medical} that mainly rely on the Bboxes or points as prompts for segmentation. While such methods do not need strict labels, they neglect the descriptive understanding of the image, revealing a deficiency in performing arbitrary description-based segmentation. In comparison, our method handles well in \textit{Labels}, free-form \textit{Text} prompts without losing the ability of \textit{Point} and \textit{Bbox}, as shown in the Table.~\ref{tab:salesman}.

\paragraph{Text Prompt Segmentation}
Text prompt segmentation, also referred to as expression segmentation~\citep{hu2016segmentation}, utilizes natural language expressions as input prompts for image segmentation tasks~\cite{liu2023gres}, moving beyond the traditional reliance on class label annotations~\citep{liu2021review}, such as nn-Unet~\citep{isensee2018nnu}, and Swin-unet~\citep{cao2022swin}. Early research in this area employed CNNs and RNNs for visual and textual feature extraction, which were later combined through feature fusion for segmentation~\citep{li2018referring}. The success of attention mechanisms further inspired a new line of work~\citep{shi2018key, ye2019cross}. More recently, transformer-based architectures have improved segmentation performance by using either carefully designed encoder-based feature fusion modules~\citep{feng2021encoder, yang2022lavt, kim2022restr} or decoder-based approaches~\citep{wang2022cris, luddecke2022image, ding2021vision}. Among these, \citep{zhou2023text} introduced a text-prompt mask decoder for efficient surgical instrument segmentation and ~\citep{zhao2024biomedparse} introduced BiomedParse to perform universal segmentation tasks in one model by text guidance. However, there are few existing works that have focused on free-form language segmentations, and the existing work~\citep{zhao2024biomedparse} spends much labor on text annotations, we provide a more efficient domain-specific solution by using retrieval augmented generation in this work. 
\paragraph{Equivariant Medical Imaging}
Equivariant neural networks ensure that their features maintain specific transformation characteristics when the input undergoes transformations, and they have achieved significant success in various image processing tasks \citep{Cohen2016Group, weiler2019e2cnn, cohen2019general, bronstein2021geometric}. Recently, equivariant networks have also been applied to medical imaging tasks, including classification \citep{winkels20183d}, segmentation \citep{kuipers2023regular, elaldi20243, he2021group}, reconstruction \citep{chen2021equivariant}, and registration \citep{billot2024se}. Equivariance can be incorporated in different ways, such as through parameter sharing \citep{finzi2021practical}, canonicalization \citep{kaba2022equivariance}, and frame averaging \citep{puny2021frame}. In our work, since we leverage a pretrained segmentation network, we achieve equivariance/invariance through canonicalization \citep{mondal2023equivariant}, which, unlike other methods, does not impose architectural constraints on the prediction network. This paper is the first to leverage the Equivariant methods to align the orientation feature in the image with the text description and achieve good medical image segmentation performance by resolving the ambiguities.

\section{Methodology}

In this section, we introduce a paradigm to equip the segmentation model with free-form language understanding ability while maintaining high segmentation accuracy. It employs the 
RAG framework to generate text prompts based on real world clinical diagnosis records. The generated free-form queries, anchored on the corresponding organ labels, are used to train a text encoder capable of efficiently interpreting the segmentation intentions (e.g., different interested organs disclosed in anatomy-informed or anatomy-agnostic prompts) and guiding the segmentation network. We also incorporate a canonicalization module, which can transform input images with arbitrary orientations into a learned canonical frame, allowing the model to produce consistent predictions regardless of the input image orientation. 

\paragraph{Preliminaries of SAM Architecture}
SAM~\citep{kirillov2023segment} contains three main parts: (1) an image encoder that transforms images into image embeddings; (2) a prompt encoder that generates prompt embeddings; (3) a mask decoder
that outputs the expected segmentation mask based on the image and prompt embeddings. Given a corresponding input medical image $x \in \mathcal{X}$ and a relevant \underline{p}rompt $p \in \mathcal{P}_x$. The image encoder embeds $x$ into $z_x$ that $z_x = \text{Encoder}^{\mathcal{X}} (x)$, similarly the prompt embedding $z_p = \text{Encoder}^\mathcal{P} (p)$. The mask decoder predicts the segmentation result (\underline{m}ask) by $ \hat{m}_x^p = \text{Decoder} (z_x, z_p)$. While the SAM model provides $\text{Encoder}^\mathcal{P}$ for spatial prompts (e.g. Bbox or point), the integration of text-based prompts has been less explored.  In text-based medical images segmentation, natural language prompts require specialized learning to effectively capture clinical terminology and segmentation intent.
\begin{figure}[t!]
  \centering
  \includegraphics[width=0.46\textwidth]{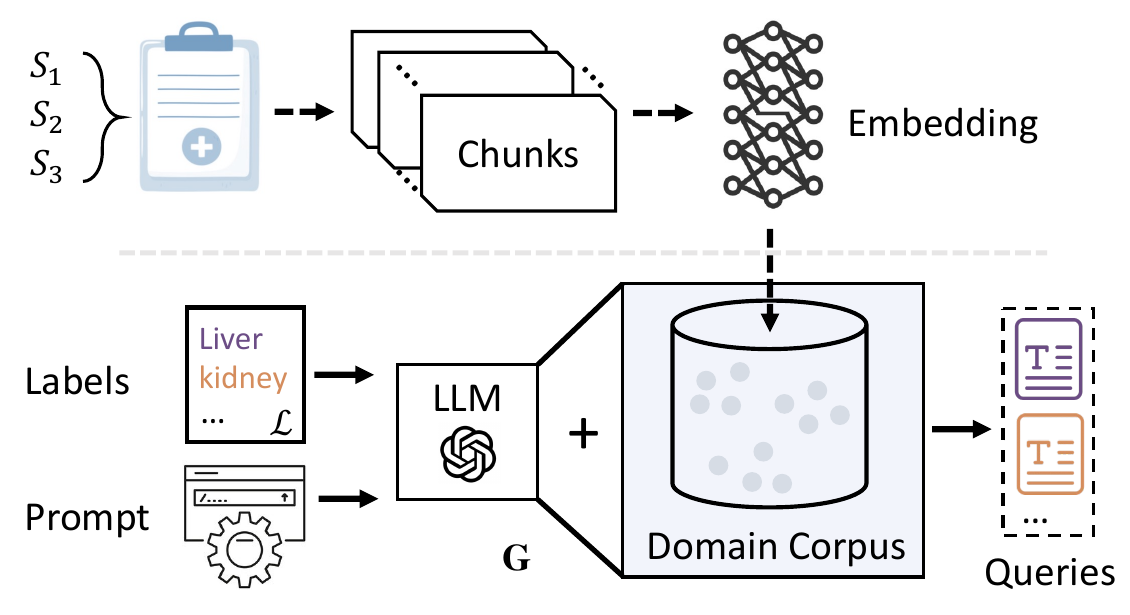}
  
  \caption{The RAG Free-form Query Generator. The domain corpus, from the EMRs embedding, completes the retrieval augmentation and enhances the LLMs with the clinical way of query.}
  \label{fig:raggenerator}
\end{figure}
\subsection{The Retrieval Augmented Query Generator}\label{sec:data}
\begin{figure*}[t!]
    \centering
    \includegraphics[width=0.98\textwidth]{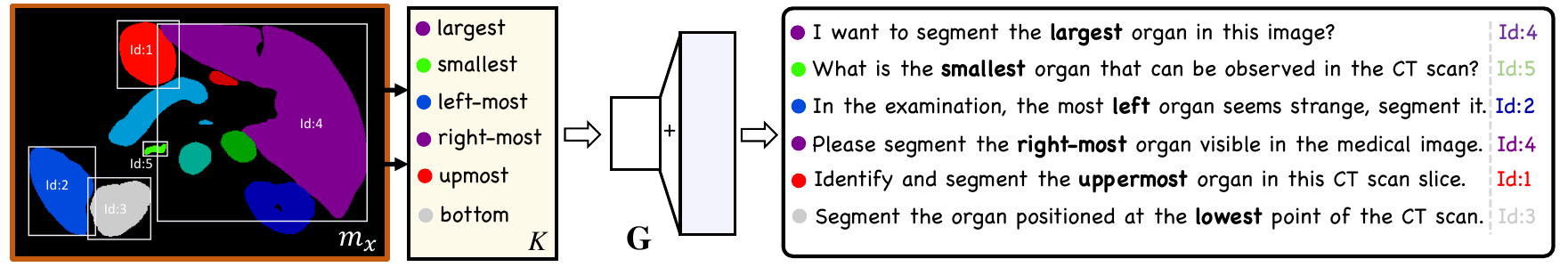}
    \caption{Spatial features extracted from the Bboxes of ground truth masks are processed by the RAG query generator $\mathbf{G}$ to produce anatomy-agnostic queries. For sample \( x \), spatial info. is extracted for mask \( m_x^{(i)} \) using Bboxes, classifying to spatial categories: largest, smallest, left-most, right-most, etc. The RAG generator \( \mathbf{G} \) then converts categories into anatomy-agnostic text queries to augment \( \mathcal{P}_x \).}
    \label{fig:query-agnostic}
\end{figure*}
\paragraph{Anatomy-Informed Query}
To equip an MIS model $\mathcal{M}$ with language comprehension abilities, it is essential to prepare a suitable natural language query \footnote{Throughout the paper, we use the terms ``query" and ``prompt" interchangeably.} corpus $\mathcal{C}$ in correspondence with the target organ label set $\mathcal{L} = \{l_1, l_2, ...l_n\}$, where $l_1 = \textit{Liver}$, $l_2 = \textit{Kidney}$, etc., as in Appendix Fig.~\ref{fig:data-statistics}. Since manual annotation is time-consuming and can be biased towards individual linguistic habits, we designed a RAG-based free-form text prompts generator to automate this process. RAG allows pre-trained LLMs to retain their free-form language generation capabilities while incorporating domain-specific knowledge and style from the provided data source $\mathcal{S}$. We collect corpus from three types of data sources. Two of these,  $\mathcal{S}_1 = \textit{Domain Expert}$, $\mathcal{S}_2 = \textit{Non-Expert}$, serve as the corpus set to simulate various styles of descriptions for segmentation purposes,. The third source, $\mathcal{S}_3 = \textit{Synthetic}$, is directly generated by GPT-4o to imitate descriptions for segmentation purposes.

For $\mathcal{S}_1 = \textit{Domain Expert}$, we collected over 7,000 reports written by doctors and identified 4,990 clinical diagnosis records that are relevant to 24 labeled organs for this study.  After de-identification, we embed such Electronic Medical Records (EMRs) into semantic vector space through Med-BERT \citep{rasmy2021med}, which outperforms the general language embedding models such as Bert or GPTs in the bioinformatics context understandings. 
Then, we built a retrieval augmented generation fashion generator agent $\mathbf{G}$, as shown in Fig.~\ref{fig:raggenerator}, provided with medical domain corpus and practitioner's language usage preference. It retains the original LLM's natural language ability such as sentences extension and rephrasing. Finally, we construct a query prompt template: ``\textit{\textbf{System:} You are an agent able to query for segmenting label \{\textcolor{custompurple}{Liver}\} in this \{CT\} scan. Please write the query sentence and output it.}" Given a label $l_i$ = \textit{Liver}, where $l \in \mathcal{L}$ regarding an arbitrary organ label with \textit{CT} modality, the $\mathbf{G}$ produces a free-form query $q_l^i$, this query is taken as prompt in the later text-aware segment model training. E.g., \textcolor{custompurple}{``(1) \textit{Examine this CT scan to determine the extent of hepatic damage present}. (2) \textit{As the symptoms suggest cirrhosis, we should analyze the related part in this CT scan for any signs of the disease}"}. These retrieved augmented results show that the interested organ may not always be explicitly mentioned, but can be inferred based on terms like `cirrhosis' and `hepatic', which are all liver-specific illnesses in clinical practice.

For $\mathcal{S}_2 = \textit{Non-Expert}$, we collected queries from people without medical training who lack knowledge of the anatomy structures to formulate the segmentation queries. For $\mathcal{S}_3 = \textit{Synthetic}$, the corpus is directly generated by LLMs. Both $\mathcal{S}_2$ and $\mathcal{S}_3$ are combined with $\mathcal{S}_1$ and processed by $\mathbf{G}$ to produce diverse and rich expression text queries for any given organ.


\paragraph{Anatomy-Agnostic Query} Anatomy-agnostic queries are crucial for training models to handle more plain descriptions (i.e., positions, sizes) that lack explicit organ names or related anatomy information. To align the anatomy-agnostic queries, $\mathcal{Q}$, with training images and their ground truth masks, we follow the process shown in Fig.~\ref{fig:query-agnostic}. Given a training sample $x$, we first retrieve spatial information for each of its mask $m_x^{(i)}$ using Bboxes, deriving spatial categories based on their positions and sizes, $k \in {\mathcal{K}}$, where the set $\mathcal{K} = \{k^{1*}, k^{2*}, \dots, k^{6*}\}$ represents six categories: largest, smallest, left-most, right-most, upmost, and bottom. The RAG generator $\mathbf{G}$ then extends this information into full language descriptions for the masks that belong to one of these six categories, generating anatomy-agnostic text queries to augment $\mathcal{P}_x$ for each $x \in \mathcal{X}$.
This pipeline, as Fig.~\ref{fig:query-agnostic}, ensures sufficient anatomy-agnostic queries are provided to train the model to segment the accurate organ masks without needing to know the organ label names.



\begin{figure*}
    \centering
    \includegraphics[width=0.95\linewidth]{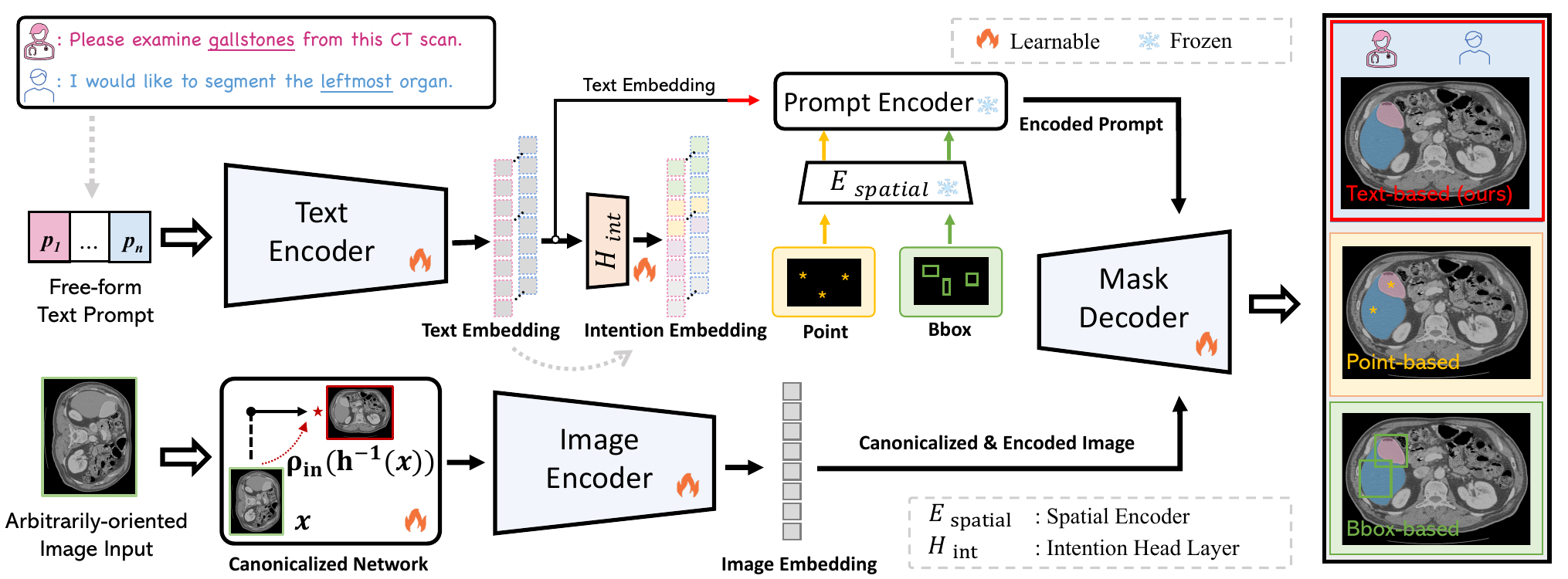}
    \caption{The architecture of our proposed model \textbf{\ours}. First, given a set of free-form text prompts $p_1 ..., p_n$, the text encoder gets the text embedding, and then passes through the learned \textit{Intention Head Layer} that maps the embedding to a space with explicit intention probabilities, which is useful for the \textbf{\ours} model weight updating as in Eq.~\ref{eq:decode}. Second, we have trained a \textit{Canonicalized Network} that transforms any medical image with arbitrary orientation into a canonicalization space, making sure the encoded image aligns with the standard clinical practice to avoid ambiguity. Third, the encoded prompts (either spatial info such as \textcolor{yellow}{Point}, \textcolor{green}{Bbox}, or \textcolor{red}{Free-form text data}), together with the encoded image, will be processed with mask decoder and output the expected masks.}
    \label{fig:enter-label}
\end{figure*}

\subsection{Free-Form Language Segmentation}
After generating a large corpus of free-form text queries via our retrieval augmented query generator, the next step is to align these queries with medical imaging segmentaion tasks. 
\paragraph{Anatomy-Informed Segmentation}\label{sec:anatomyinform}
 For anatomy informed text prompts, the text encoder must learn embeddings that group similar organ segmentation intents together while clearly separating unrelated intents in distinct semantic clusters. 
We adopt the CLIP~\citep{radford2021learning} as the foundation of text encoder for its capability of understanding semantics. Given a text prompt $p \in \mathcal{P}_x$ associated with the image $x$, 
the CLIP text encoder converts it into an embedding vector $\mathbf{t}_p$ in a shared embedding space: $\mathbf{t}_p = \text{Encoder}^{\mathcal{P}}(p) \in \mathbb{R}^{D}$, where $D$ is the dimensionality of the text embedding space. 

To further strengthen the model’s ability to differentiate between organ segmentation, we introduce an intention head on top of the text embeddings by CLIP. This head is a linear layer $\mathbf{W}_{\text{cls}} \in \mathbb{R}^{C \times D}$, where $C = 24$ is the number of organ class. The intention logits $\mathbf{y}_p$ are derived for each encoded vector $\mathbf{t}_p$:  $\mathbf{y}_p = \mathbf{W}_{\text{cls}} \mathbf{t}_p + \mathbf{b}_{\text{cls}}$. Given a corresponding medical image embedding $z^x$, we train the model by following loss function:
\begin{equation}\label{eq:decode}
    L = \mathop{\arg\min}_{\mathcal{W}^*} 
     \frac{1}{|\mathcal{X}|} \sum_{x \in \mathcal{X}} \frac{1}{|\mathcal{P}_x|} \sum_{p \in \mathcal{P}_x} 
    \big[\mathcal{L}_{\text{Dice}}(\hat{m}_x^p, m_x^p) + \mathcal{L}_{\text{ce}^*}\big]
\end{equation} 
where $\mathcal{W}^*$ are sets of model weights: $\mathcal{W}^* = \{ \mathbf{W}_\text{cls}, \mathbf{b}_{\text{cls}},  \mathbf{W}^{E}, \mathbf{W}^{D}, \mathbf{W}^{P}\}$, and $\hat{m}_x^p = \text{Decoder} (z_x, \mathbf{t}_p)$ and $m_x^p$ are predicted and ground truth masks. $l_p \in [0, ..., 23]$ is the ground truth organ class for the prompt. 
$\mathbf{W}^{E}$, $\mathbf{W}^{D}$ and $\mathbf{W}^{P}$ represent the image encoder, decoder and CLIP text encoder weights, respectively. We use both Dice loss $\mathcal{L}_{\text{Dice}}$ and cross-entropy loss $\mathcal{L}_{\text{ce}^*}$ for predicted masks. And for $\mathcal{L}_{\text{ce}^*}$:

\begin{equation}
    \mathcal{L}_{\text{ce}^*} = \mathcal{L}_{\text{ce}}(\hat{m}_x^p, m_x^p) + \mathcal{L}_{\text{ce}}(\mathbf{y}_p, l_p).
\end{equation}

it contains two parts, apart from measuring the predicted mask $\hat{m}_x^{p}$ compared to the ground truth mask $m_x^{p}$, the second term $\mathcal{L}_{\text{ce}}(\mathbf{y}_p, l_p)$ encourages the model to correctly classify organs based on text prompts, ensuring the text embedding aligns with the intended organ class.

\paragraph{Anatomy-Agnostic Segmentation}
For anatomy-agnostic descriptions, which do not explicitly mention specific organs but instead focus on spatial attributes (e.g., ``leftmost", ``largest"), the model must learn from spatial features $k_x \in \mathcal{K}$ to pair with the corresponding mask $m_x^k$ for every $x \in \mathcal{X}$. Anatomy-agnostic queries share the same embedding space as anatomy-informed queries, but $k_x$ is not necessarily associated with a specific organ. In this case, we use the same loss function as shown in Eq. \ref{eq:decode} but without the last classification term. 


\newcolumntype{P}[1]{>{\centering\arraybackslash}p{#1}}
\begin{table*}[h!]
\small
    \centering
    \caption{\textbf{Anatomy-Informed} Segmentation Results: \ours{} consistently outperforms baselines on both organ name and free-form text prompts segmentation tasks, demonstrating superior language understanding and segmentation accuracy across in-domain and out-of-domain datasets, even when applied with random transformations.}
    \label{tab:anatomy_informed}
    \begin{tabular}{P{2.6cm}|P{0.7cm}P{0.7cm}|P{0.7cm}P{0.7cm}|P{0.7cm}P{0.7cm}|P{0.7cm}P{0.7cm}|P{0.7cm}P{0.7cm}|P{0.7cm}P{0.7cm}}\toprule
        \multirow{2}{*}{\textbf{Organ Name}}   & \multicolumn{2}{c|}{FLARE} & \multicolumn{2}{c|}{WORD} & \multicolumn{2}{c|}{RAOS} & \multicolumn{2}{c|}{TransFLARE}& \multicolumn{2}{c|}{TransWORD} & \multicolumn{2}{c}{TransRAOS}\\ 
        \cmidrule(lr){2-3} \cmidrule(lr){4-5} \cmidrule(lr){6-7} \cmidrule(lr){8-9} \cmidrule(lr){10-11} \cmidrule(lr){12-13} 
         &  Dice &  NSD & Dice &  NSD &   Dice & NSD &  Dice &  NSD &  Dice &  NSD & Dice &  NSD \\ \midrule
         CLIP+MedSAM & 0.473 & 0.518 & 0.411 & 0.446 & 0.475 & 0.440  & 0.388&0.417&0.357& 0.437 & 0.352 & 0.399\\
         MedCLIP+MedSAM &  0.557 & 0.516 & 0.466 &  0.510 & 0.419 & 0.320 & 0.485 & 0.415 & 0.342 & 0.378 & 0.336 & 0.336\\
         Universal Model & 0.649 & 0.697& 0.512& 0.408 & 0.442 & 0.301& 0.380 & 0.290& 0.299& 0.278 & 0.200 & 0.201\\
         UniverSeg & 0.790 & 0.777& 0.722 & 0.739 & 0.719 & 0.754 & 0.160 & 0.190& 0.413 & 0.429 & 0.322 & 0.318\\
         BiomedParse & 0.852 & 0.913 & 0.738 & 0.815 & 0.708 & 0.769 & 0.824 & 0.894 & 0.648 & 0.723  & 0.592  & 0.649 \\
         \textbf{\ours{}} & \textbf{0.908} & \textbf{0.956} & \textbf{0.837} & \textbf{0.884} & \textbf{0.852} & \textbf{0.883}  & \textbf{0.898} & \textbf{0.949} & \textbf{0.835} & \textbf{0.875} & \textbf{0.847} & \textbf{0.879}\\ 
        \bottomrule
    \end{tabular}
    \begin{tabular}{P{2.6cm}|P{0.7cm}P{0.7cm}|P{0.7cm}P{0.7cm}|P{0.7cm}P{0.7cm}|P{0.7cm}P{0.7cm}|P{0.7cm}P{0.7cm}|P{0.7cm}P{0.7cm}}\toprule
        \multirow{2}{*}{\textbf{Free Form}}   & \multicolumn{2}{c|}{FLARE} & \multicolumn{2}{c|}{WORD} & \multicolumn{2}{c|}{RAOS} & \multicolumn{2}{c|}{TransFLARE}& \multicolumn{2}{c|}{TransWORD} & \multicolumn{2}{c}{TransRAOS}\\ 
        \cmidrule(lr){2-3} \cmidrule(lr){4-5} \cmidrule(lr){6-7} \cmidrule(lr){8-9} \cmidrule(lr){10-11} \cmidrule(lr){12-13} 
         &  Dice &  NSD & Dice &  NSD &   Dice & NSD &  Dice &  NSD &  Dice &  NSD & Dice &  NSD \\ \midrule
         CLIP+MedSAM & 0.425 & 0.468 & 0.381& 0.347& 0.402& 0.400 & 0.342 & 0.434 & 0.356& 0.456 & 0.339& 0.357\\
         MedCLIP+MedSAM & 0.696 & 0.557 & 0.473 & 0.518 & 0.365 & 0.424&0.483&0.501&0.239& 0.241 & 0.307 & 0.331\\
         Universal Model & --- & --- & --- & --- & --- & --- & --- & --- & ---  & ---  & ---  & --- \\
         UniverSeg & --- & --- & --- & --- & --- & --- & --- & --- & ---  & ---  & ---  & --- \\
         BiomedParse & 0.817 & 0.877 & 0.782 & 0.864 & 0.754 & 0.821 & 0.829 & 0.901 & 0.730  & 0.811  & 0.702  & 0.767 \\
         \textbf{\ours{}} & \textbf{0.912} & \textbf{0.958} & \textbf{0.830} & \textbf{0.889} & \textbf{0.854} & \textbf{0.885} & \textbf{0.896} & \textbf{0.942} & \textbf{0.833} & \textbf{0.888} & \textbf{0.865} & \textbf{0.899} \\ 
        \bottomrule
    \end{tabular}

\end{table*}

\subsection{Semantics-Aware Canonicalization Learning}\label{sec:cano}

We incorporate roto-reflection symmetry \citep{Cohen2016Group} into our architecture for two key reasons: 1) Organs and anatomical structures can appear in various orientations and positions due to differences in patient positioning, imaging techniques, or inherent anatomical variations. Equivariance ensures that the model's segmentation adapts predictably to transformations of the input image. 2) We aim to ensure our model reliably interprets and segments organs that have positional terms in their names, such as ``left" or ``right kidney" from text prompts, regardless of the scan's orientation, thereby enhancing the model's robustness and accuracy.

Following \cite{kaba2022equivariance, mondal2023equivariant}, we train a separate canonicalization network $ {h}: \mathcal{X} \mapsto G$, where $\mathcal{X} $ represents the medical image sample space, $G$ represents the desired group, and $h$ is equivariant to $G$. This network generates group elements that transform input images into canonical frames, standardizing the image orientation before applying the prediction function. 
\begin{equation}\label{eq:can}
    f(x) = \rho_{\text{out}}({h}(x)) \; p( \rho_{\text{in}}({h}^{-1}(x))x,\;\textbf{t})
\end{equation}

The Eq.~\ref{eq:can} shows how this canonicalization process maps the transformed input back to a common space where the segmentation prediction network $p$ operates, where $p$ is the segmentation prediction network (composed of the Image Encoder and Mask Decoder in Figure \ref{fig:enter-label}), $\textbf{t}$ is the text prompt embedding produced by our text encoder, and  $\rho_{\text{in}}$ and $\rho_{\text{out}}$ are input and output representations. The segmented images or masks produced by $p$ can be transformed back with $\rho_{\text{out}}({h}(x))$ as needed. Without this transformation, $f$ is invariant; otherwise, it is equivariant. Thus, the \ours{} architecture visualized in Figure \ref{fig:enter-label} is invariant. We use ESCNN \citep{cesa2022a} to build the canonicalization network. This approach has the advantage of removing the constraint from the main prediction network and placing it on the network that learns the canonicalization function. Appendix \ref{app:symmetry} provides a detailed introduction of symmetry and equivariant networks.

As the entire architecture achieves invariance or equivariance through canonicalization, the model produces the same segmentation or consistently transforms the segmentation according to the transformed input. In other words, the model always segments the same areas of interest regardless of the image's orientation with the same text prompt. For example, as long as the ground truth  ``right kidney" mask of a CT image has been shown to the model once, no matter how the orientation of the  CT image and the location of the right kidney change, the model will always segment the same area.

However, without proper training, ${h}(x)$ might map different images to inconsistent canonical frames, causing a distribution shift in the inputs to the prediction network and affecting performance. Thus, training the canonicalization network together with the segmentation prediction network is essential to ensure consistent mapping to the desired frame. It is worth noting that users can choose to disable the canonicalizer when working with anatomy-agnostic prompts, as the segmented organ may differ if the original image is not in the canonical frame. The decision depends on whether the user wants to segment the original or the canonicalized image, as the model will segment whatever image is fed into the image encoder based on the provided text prompts.
\subsection{Training Strategy}

We employ a three-stage training strategy for \ours{}: 1) \textbf{Learning canonicalization}: The canonicalization network is trained independently using FLARE22 training samples with random transformations from the O(2) group. It is optimized with MSE loss between the canonicalized samples and their original counterparts, encouraging the network to map transformed samples back to their canonical orientations in the FLARE22 dataset and avoid arbitrary orientations that could impair prediction performance. 2) \textbf{Learning text-prompted segmentation}: We train \ours{} with queries from Generator $\mathbf{G}$ as introduced in Section~\ref{sec:data}, using both anatomy-informed and anatomy-agnostic prompts on original scans, ensuring the segmentation network accurately responds to different prompts without interference from canonicalization. 3) \textbf{Learning augmentation and alignment}: In the final stage, we perform joint training on all scans with random O(2) transformations. As the canonicalization network may not always provide the exact canonical orientation expected by the segmentation network initially, this serves as free augmentation, gradually aligning the canonicalization and segmentation networks.

\section{Experiment}
\subsection{Datasets and Experiments Setup}



\textbf{Image Datasets}: We collected CT scans covering 24 labeled organs from the datasets below. Of these, 1,089 scans from MSD~\citep{antonelli2022medical}, BTCV~\citep{gibson_2018_1169361}, WORD~\citep{luo2021word}, AbdomenCT-1K~\citep{Ma-2021-AbdomenCT-1K}, FLARE22~\citep{FLARE22}, and CHAOS~\citep{CHAOSdata2019} were used for training, while 65 scans from FLARE22, WORD, and the RAOS~\citep{luo2024rethinking} test set were used for evaluation. Pre-processing steps (slice filtering, intensity scaling) standardized quality and reduced domain gaps, \textbf{resulting in 91,344 images for training/validation and 9,873 for testing}. Details on dataset statistics and pre-processing steps are provided in the Appendix~\ref{appendix:first}. \textbf{Text Datasets}: Our text dataset includes 12,120 training queries (600 anatomy-agnostic, 11,520 anatomy-informed) generated by the RAG query generator. For testing, we created 2,880 queries (120 per organ) covering both in-domain and out-of-domain queries. \textbf{Experiment Setups}: All experiments were conducted on AWS instance equipped with 8 V100 GPUs, each with 32 GB of memory, more details in Appendix~\ref{appendix:first} and ~\ref{appendix:train}. 


\begin{figure*}
    \centering
    \includegraphics[width=0.99\linewidth]{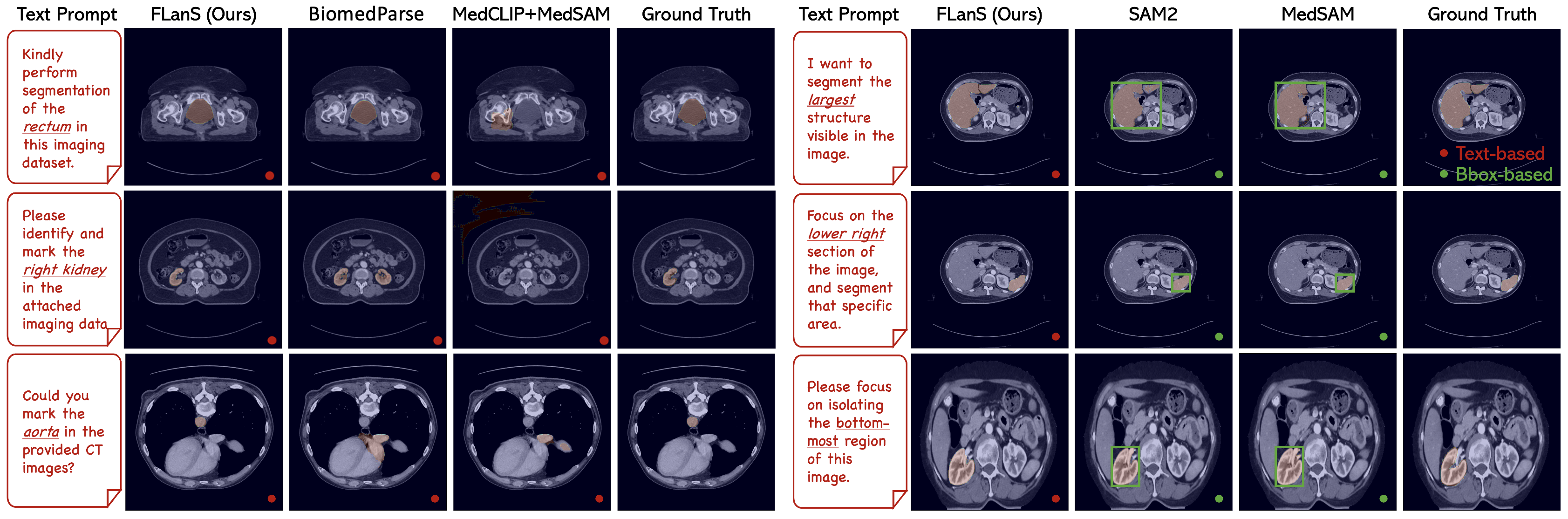}
    \caption{\textbf{Left}: Segmentation with anatomy-\textbf{informed} prompts. We could observe that \ours can precisely segment the organ described in free-form text prompts, while other baselines make mistakes in identifying the organs. \textbf{Right}: Segmentation with anatomy-\textbf{agnostic} prompts. We could observe that the \ours is texture-aware, descriptions of the sizes and positions can be understood, and is competitively accurate to the direct Bbox segment.}
    \label{fig:FLans}
\end{figure*}

\begin{figure*}[htb!]
  \begin{minipage}[b]{0.28\textwidth}
   \includegraphics[width=\textwidth]{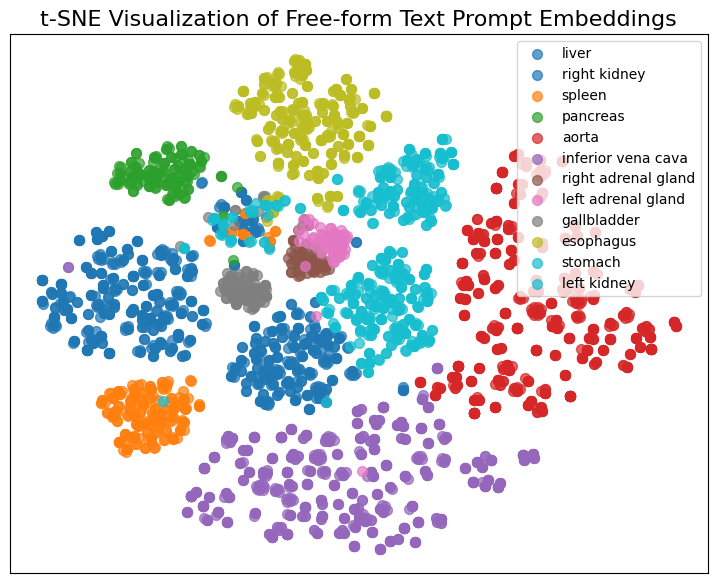}
  \end{minipage} \hspace{5pt}
   \begin{minipage}[b]{0.69\textwidth}
   \includegraphics[width=\textwidth]{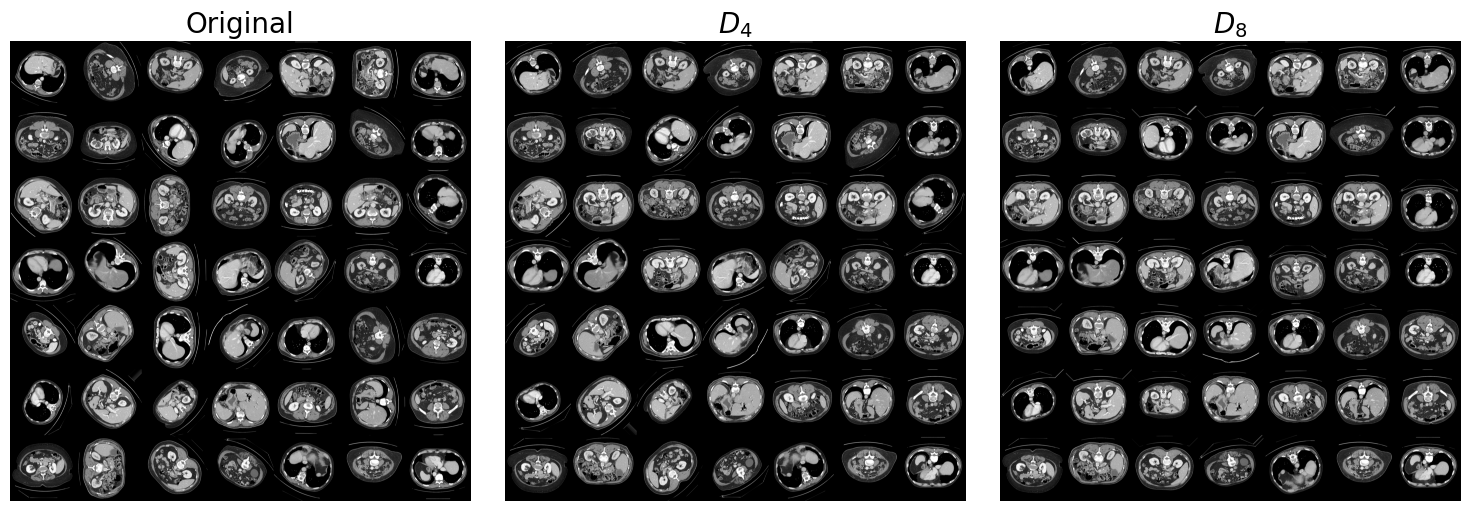}
  \end{minipage} 
\caption{\textbf{Left}: t-SNE visualization of the free-form text prompt embedding. \ours (Ours) can effectively distinguish between different organ-related queries. \textbf{Right}: Canonicalized CT scans from $D_4$ and $D_8$ canonicalization networks for a batch of randomly transformed scans from the FLARE22 dataset. Medical images can be successfully transformed back to an aligned canonicalization space.}
\label{fig:ablation_study}
\end{figure*}

\subsection{Anatomy-Informed Segmentation}

We first compare our model, \ours{}, with the SOTA baselines on a held-out subset of the FLARE22 training set (FLARE), the public WORD validation set (WORD), and RAOs cancer CT images (RAOS). Both FLARE22 and WORD serve as in-domain test sets, while RAOS is an out-of-domain test set, as neither our model nor the baselines were trained on this dataset. Although the original test sets already contain scans with varying orientations, we further evaluated the models’ robustness by applying random O(2) transformations to the three test sets, creating additional sets: TransFLARE, TransWORD, and TransRAOS. More importantly, we tested the models using Anatomy-Informed text prompts, which included two types: purely organ names and free-form text descriptions. 


Among existing baselines, BiomedParse \citep{zhao2024biomedparse} is, to our best knowledge, the only model designed to handle free-form text prompts. While \texttt{Universal Model} \citep{liu2023clip} is also a medical imaging foundation model that incorporates text descriptions during training, it performs segmentation at the inference stage using organ IDs. Consequently, we evaluate this model with prompts consisting solely of organ names. Another widely used approach for text-prompt segmentation involves combining CLIP-based models \citep{radford2021learning} with segmentation models \citep{li2024clipsam, wang2024samclip}. In these methods, segmentation models first generate potential masks based on a set of random bounding box or point prompts that span the entire image. CLIP-based models then embed both the text prompt and the cropped images from these masks. The final mask is selected based on the highest similarity between the cropped image embedding and the text embedding. To cover this approach, we include two additional baselines: 1) \texttt{CLIP + MedSAM}, where MedSAM \citep{wu2023medical} is SAM \citep{kirillov2023segment} fine-tuned on medical imaging datasets; and 2) \texttt{MedCLIP + MedSAM}, where MedCLIP \citep{wang2022medclip}, a contrastive learning framework trained on diverse medical image-text datasets, is paired with MedSAM for segmentation. We include a SOTA few-shot learning model, \texttt{UniverSeg} \citep{butoi2023universeg}, to ensure our baselines remain competitive. Unlike the other methods, \texttt{UniverSeg} does not support text prompts. Instead, we evaluate it using a few support samples with their corresponding ground truth organ masks, rather than prompts based on organ names.

As we can see from Table \ref{tab:anatomy_informed}, \ours{} achieves superior performance in segmenting based on organ name. More importantly, \ours significantly outperforms the baselines on free-form text prompts segmentation, where the baselines struggle with more complex language input. This suggests that training with diverse free-form text prompts enhances the model's ability to understand language and the relationship between text descriptions and medical images. Furthermore, \ours{} maintains high Dice and NSD scores on the transformed test sets thanks to the help of the canonicalization network. The left panel of Fig.~\ref{fig:FLans} visualizes the segmentations generated by the best baseline and \ours{}, alongside their corresponding text prompts, illustrating our model’s superior language understanding and segmentation accuracy.

\subsection{Anatomy-Agnostic Segmentation}

\begin{table}
\small
    \centering
    \caption{\textbf{Anatomy Agnostic} Segmentation Results: Comparison of \ours{} using positional and size information text prompts vs. \texttt{MedSAM} and \texttt{SAM2} using Bboxes or points. \ours{} achieves competitive or superior performance across both in-domain and out-of-domain test sets.}
    \label{tab:anatomy_agnostic}
    \begin{tabular}{P{2.7cm}|P{0.5cm}P{0.6cm}|P{0.5cm}P{0.6cm}|P{0.5cm}P{0.6cm}}\toprule
        \multirow{2}{*}{\textbf{Methods}}   & \multicolumn{2}{c|}{FLARE} & \multicolumn{2}{c|}{WORD} & \multicolumn{2}{c}{RAOS (\textit{OOD})}  \\ 
        \cmidrule(lr){2-3} \cmidrule(lr){4-5} \cmidrule(lr){6-7} 
         &  Dice &  NSD  & Dice &  NSD  &  Dice &  NSD   \\ \midrule
         SAM2-large (\textit{Point}) & 0.585 & 0.652 & 0.534 & 0.551 &  0.488& 0.497\\
         SAM2-large (\textit{Bbox}) & 0.873 & \textbf{0.906} & 0.848 & 0.802 & 0.818 & 0.749\\
         MedSAM (\textit{Bbox}) & \textbf{0.887} & 0.872 & 0.783 & 0.781 & 0.697 & 0.681\\ 
         \textbf{\ours{}} \textit{(Free-form)} & 0.844 & 0.841 & \textbf{0.855} & \textbf{0.853} & \textbf{0.851} & \textbf{0.850}\\ 
        \bottomrule
    \end{tabular}
\end{table}

To evaluate our model’s ability to understand anatomy-agnostic text prompts, we tested its segmentation performance using prompts that contain solely positional or size-related information. To the best of our knowledge, no existing model is designed to handle anatomy-agnostic text prompts. Therefore, we chose state-of-the-art \texttt{MedSAM} \citep{wu2023medical} (SAM fine-tuned on medical imaging datasets) and the latest \texttt{SAM2} \citep{ravi2024sam2} as baselines. However, instead of text prompts, these models were provided with ground-truth organ Bboxes or point prompts. Our goal in this experiment is for \ours{} to achieve comparable results to the baselines because \ours{} is only given text prompts with positional or size information while the baselines are given the bounding box or point prompts of ground truth organ. 

As shown in Table \ref{tab:anatomy_agnostic},  achieves top performance across in-domain and out-of-domain test sets. \texttt{MedSAM} performs well on the FLARE and WORD test sets but struggles on the RAOS due to the lack of training on that dataset. \texttt{SAM2}, when provided with bounding box prompts, consistently performs well across all test sets and shows strong generalizability. However, its performance degrades with point prompts, likely because medical scans lack the distinct edges present in the datasets \texttt{SAM2} was originally trained on. The right panel of Fig.~\ref{fig:FLans} visualizes the segmentations produced by the best baseline and \ours{}, along with anatomy-agnostic text prompts. \ours{} reliably segments the correct organs based on the provided positional or size descriptions.

\subsection{Ablation Study on the Model Architecture}

\begin{table}[htbp]
    \centering
    \small
    \caption{Ablation study: prediction performance of \ours and its variants with progressively removed components on the FLARE22 original and transformed test sets. Each row represents a version of the model with one additional component removed.}
    \label{tab:ablation_study}
    \resizebox{0.44\textwidth}{!}{
    \begin{tabular}{P{2.4cm}|P{0.9cm}P{1.1cm}|P{0.9cm}P{1.1cm}}
        \toprule
        \multirow{2}{*}{Model Variants}   & \multicolumn{2}{c|}{Canonicalized Test Set} & \multicolumn{2}{c}{Transformed Test Set}\\ 
        \cmidrule(lr){2-3} \cmidrule(lr){4-5} 
        &  Dice &  NSD & Dice &  NSD \\ \midrule
        \textbf{\ours{}} (full model) & \textbf{0.901\tiny{$\pm$0.003}} & \textbf{0.953\tiny{$\pm$0.008}} & \textbf{0.895\tiny{$\pm$0.010}} & \textbf{0.951\tiny{$\pm$0.002}}  \\
        – Canonicalization & 0.865\tiny{$\pm$0.010} & 0.896\tiny{$\pm$0.011}  & 0.685\tiny{$\pm$0.012}  & 0.728\tiny{$\pm$0.014}   \\
        – Data Augm  & 0.883\tiny{$\pm$0.012} & 0.930\tiny{$\pm$0.017}  & 0.289\tiny{$\pm$0.011} & 0.328\tiny{$\pm$0.019} \\ 
        – Trainable ImgEnc & 0.748\tiny{$\pm$0.009} & 0.845\tiny{$\pm$0.016} & 0.301\tiny{$\pm$0.009} & 0.283\tiny{$\pm$0.017 }  \\ 
        – Classification Loss & 0.718\tiny{$\pm$0.036} & 0.831\tiny{$\pm$0.029}  & 0.271\tiny{$\pm$0.020} & 0.234\tiny{$\pm$0.049}  \\ 
        \bottomrule
    \end{tabular}
    }
\end{table}

We conducted an ablation study of \ours{} on the FLARE22 dataset \citep{FLARE22} to understand the contribution of each component, as presented in Table \ref{tab:ablation_study}. Using an 80\%-10\%-10\% train-validation-test split on the public FLARE22 training set, we evaluate the models' performance on both the held-out test set and a transformed test set, which contained samples applied with random transformations from $O(2)$. Table \ref{tab:ablation_study} shows the prediction performance of \ours{} and its variants, with components progressively removed. The results highlight that each component plays a crucial role in the model's overall performance. Notably, while data augmentation improved the model's robustness to random transformations, it slightly reduced performance on the canonical test set, as the model had to handle various transformations. However, by canonicalization network, the segmentation backbone focuses specifically on canonicalized medical images, thus achieving the best performance on both test sets.

\subsection{Understanding of Free-Form Text Prompts}
Fig.~\ref{fig:ablation_study} left visualizes the t-SNE embeddings of free-form text prompts corresponding to all 13 FLARE22 data classes, including liver, right kidney, spleen, and others. The text prompt encoder effectively clusters these prompts, revealing anatomically structured semantics. This demonstrates \ours{} has a strong capability in understanding and distinguishing free-form text prompts.


\subsection{Effectiveness of the Canonicalization}
\begin{figure}[h!]
    \centering
    \includegraphics[width=0.45\textwidth]{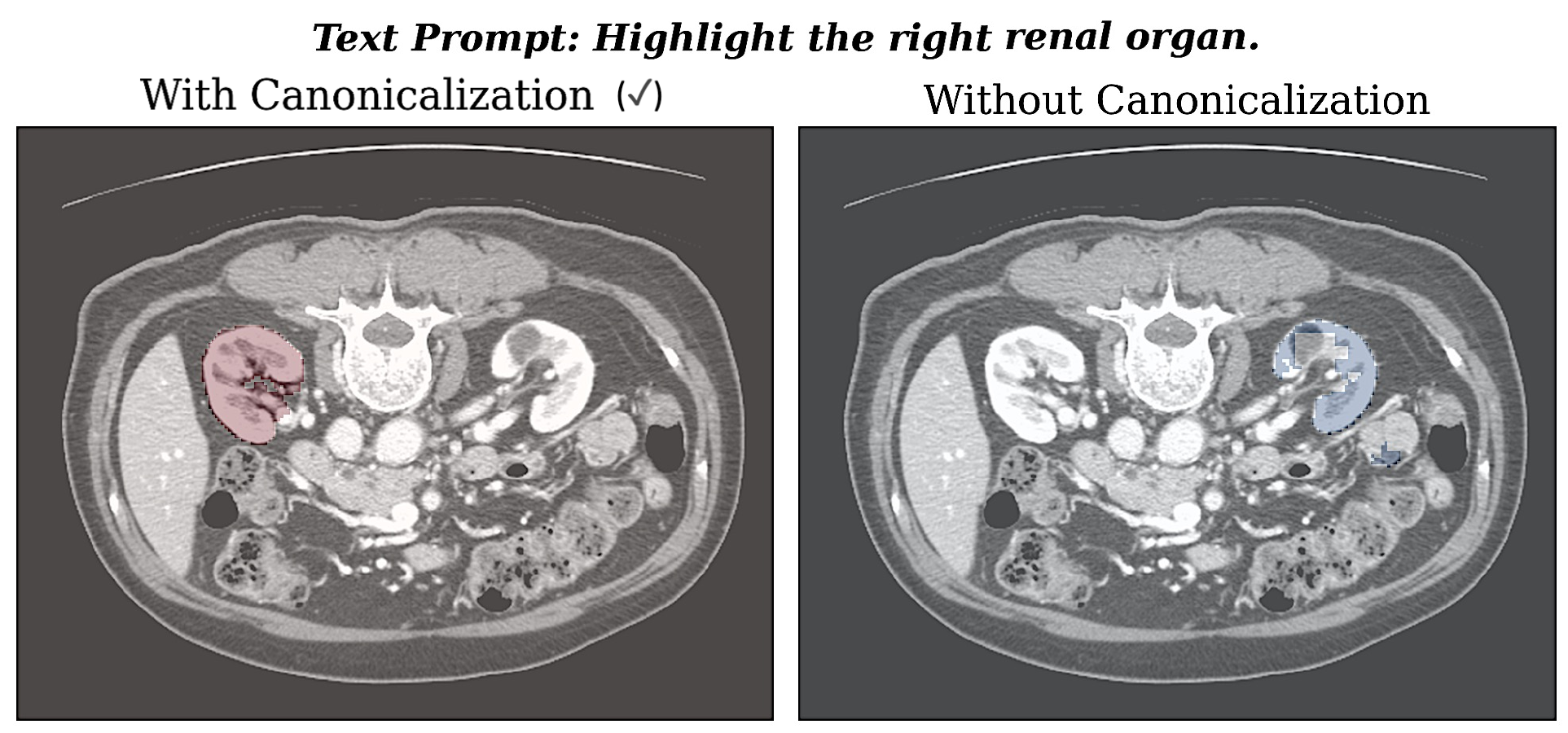}
    \caption{The model without canonicalization incorrectly highlights the left kidney due to confusion between anatomical position (``right kidney") and the organ's appearance on the right side of the image.}
    \label{fig: can_vs_baseline}
\end{figure}

\begin{figure*}[h!]
    \centering
    \includegraphics[width=0.90\linewidth]{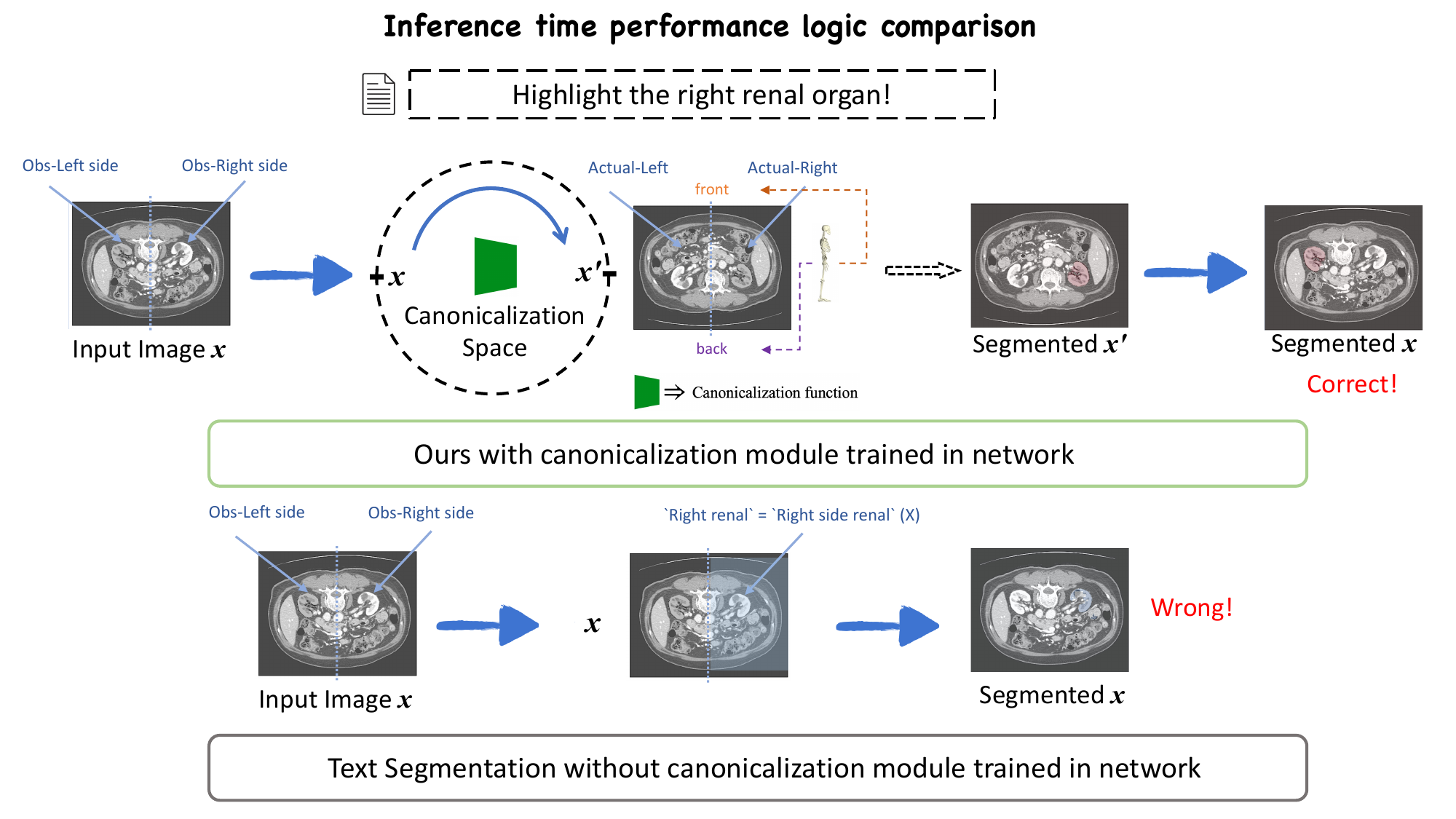}
    \caption{The explanation behind the segmentation logic of two methods shown in the figure~\ref{fig: can_vs_baseline}. \ours trained with canonicalization module, is able to distinguish that, the question asking about `right renal' is the actual `right' to the human body, not the observation side, since `right kidney' is essentially different from `right side kidney', and our model handle it by a well-learned canonicalization space and transforming the input image $x$ to the $x'$ based on the body standard orientation. The segmentation without such a canonicalization module can only directly process the input image, thus making mistakes when it comes to language ambiguity.}
    \label{fig:explains}
\end{figure*}

The right side of Fig.~\ref{fig:ablation_study} shows the canonicalized CT scans from $D_4$ and $D_8$ canonicalization networks for a batch of original scans from the FLARE22 dataset applied with random transformations from $O(2)$ group. Increasing the group order of the canonicalization network improves alignment to a canonical orientation. Using a simple three-layer network with a hidden dimension of 8 and a kernel size of 9 effectively achieves this alignment. 


\textbf{Discussion: Why not just perform augmentation but apply canonicalization?} Canonicalization ensures equivariance, resolving positional confusion that simple augmentation cannot. For instance, in  Fig.~\ref{fig: can_vs_baseline}, a model using only augmentation misidentifies the left kidney as the right one when given the prompt “Highlight the right renal organ,” because the input scan is not in the standard orientation and the right side kidney appears on the left side of the image. Canonicalization addresses this by aligning scans to a consistent orientation, enabling accurate segmentation regardless of the scan’s initial positioning. 
Specifically, \ours trained with canonicalization module, is able to distinguish that, the question asking about `right renal' is the actual `right' to the human body, not the observation side, since `right kidney' is essentially different from `right side kidney', and our model handle it by a well-learned canonicalization space and transforming the input image to the canonicalized one based on the body standard orientation and then perform the segmentation as shown in Figure.~\ref{fig:explains}. 


\section{Conclusion}

We presented \ours{}, a medical image segmentation model effectively handling diverse free-form prompts. By dividing the real-world query into anatomy-informed and anatomy-agnostic scenarios, we could successfully train a flexible foundation model by proposing a RAG-based generator for description generation and the equivariance module for efficient image feature capturing. Through this design, \ours learns a strong understanding of the free-form language descriptions that promote arbitrary segmentation across varying orientations with high accuracy. The model has the potential to generalize to more modalities and organ classes for real-world clinical deployment.

\section{Limitations and Future Work}
This paper explores a solution that bridges the language flexibility and difficult segmentation tasks in the medical image domain. Although we demonstrate that the current model performs well in CT scans with various text prompts, we found that its performance in other modalities (e.g., MRI or Ultrasound) is still limited. Besides, the current model's ability in processing positional hints is fixed to types (latest, smallest, or left-most, etc), to advance the flexibility, possible explorations can be made on combinational prompts, e.g., Size + Positional is an interesting direction.


\section*{Acknowledgments}
The work was partially supported by NSF awards \#2421839. The views and conclusions in this paper are those of the authors and should not be interpreted as representing any funding agencies.

\bibliographystyle{ACM-Reference-Format}
\balance
\bibliography{main}

\newpage
\appendix
\section{Equivariance and Symmetry}

\label{app:symmetry}
Equivariant neural networks are designed to explicitly incorporate symmetries that are present in the underlying data. Symmetries, often derived from first principles or domain knowledge, such as rotational or translational invariance, allow the network to process inputs in a way that is consistent with these transformations. This is particularly important when the ground truth functions respect such symmetries, as the incorporation of these properties can significantly enhance model performance and generalization.

\paragraph{Group}
A group of symmetries or simply \textit{group} is a set $G$ together with a binary operation $\circ \colon G \times G \to G$ called \textit{composition} satisfying three properties: 1) \textit{identity}: There is an element $1 \in G$ such that $1 \circ g = g \circ 1 = g$ for all $g \in G$; 2) \textit{associativity}: $(g_1 \circ g_2) \circ g_3 = g_1 \circ ( g_2 \circ g_3)$  for all $g_1,g_2,g_3 \in G$; 3) \textit{inverses} if $g \in G$, then there is an element $g^{-1} \in G$ such that $g \circ g^{-1} = g^{-1} \circ g = 1$.

Examples of groups include the dihedral groups $D_4$ (symmetries of a square) and $D_8$ (symmetries of an octagon), as well as the orthogonal group $O(2)$, which represents all rotations and reflections in 2D space. Both $D_4$ and $D_8$ are discrete subgroups of $O(2)$. 

\paragraph{Representation}
A group representation defines how a group action transforms elements of a vector space by mapping group elements to linear transformations on that space. More specifically, a group representation of a group $G$ on a vector space $V$ is is a homomorphism: $\rho\colon G \to \text{GL}(X)$, where $\text{GL}(X)$ is the group of invertible linear transformations on $V$. This means for any $g_1, g_2 \in G$, $\rho$ is a linear transformation (often represented by a matrix) such that the group operation in $G$ is preserved:
\begin{equation}
\rho(g_1 g_2) = \rho(g_1)\rho(g_2)
\end{equation}

\begin{figure*}[h!]
    \centering
    \includegraphics[width=0.99\linewidth]{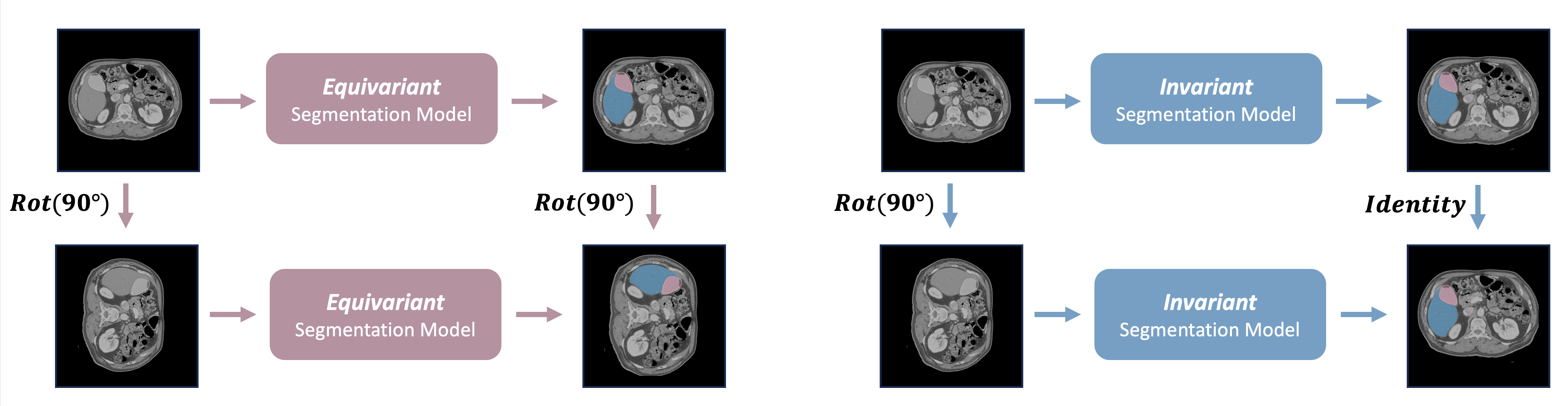}
    \caption{An equivariant model (left) ensures that its output transforms in a specific, predictable way under a group of transformations applied to the input, preserving the structure of the transformation (e.g., rotating the input results in a correspondingly rotated output). In contrast, an invariant model (right) produces an output that remains unchanged regardless of any transformations applied to the input from the same group.}
    \label{fig:symmetry}
\end{figure*}

\paragraph{Equivariance}
Formally, a neural network is said to be equivariant to a group of transformations $G$ if applying a transformation from the group to the input results in a corresponding transformation to the output. Mathematically, for a function $f \colon X \to Y$ to be \textbf{$G$-equivariant}, the following condition must hold: 
\begin{equation}
    f( \rho_{\text{in}}(g)(x)) = \rho_{\text{out}}(g) f(x) 
\end{equation}
for all $x \in X$ and $g \in G$, where $\rho_{\text{in}}\colon G \to \text{GL}(X)$ and $\rho_{\text{out}}  \colon G \to \text{GL}(Y)$ are input and output representations \citep{bronstein2021geometric}. Invariance is a special case of equivariance where the output does not change under the group action. This occurs when the output representation $\rho_{\text{out}}(g)$ is trivial. Figure \ref{fig:symmetry} visualize how the equivariant and invariant networks work. 

\paragraph{Equivariance via weight-sharing}
One of the primary approaches to incorporating symmetry into neural networks is through weight sharing \citep{satorras2021n, cohen2018general, wangdiscovering}. This approach enforces equivariance by constraining the network's architecture so that the weights are shared across different group elements. For example, in $G$-convolutions \citep{Cohen2016Group}, the same set of weights is shared across the transformed versions of the input, ensuring that the network's predictions remain consistent under those transformations. In a layer of $G$-steerable CNNs \citep{weiler2019e2cnn}, a set of equivariant kernel bases is precomputed based on the input and output representations, and the convolution kernel used is a linear combination of this equivariant kernel basis set, where the coefficients are trainable. Similar approaches can also be used to develop equivariant graph neural networks \citep{geiger2022e3nn}. These architectures directly modify the network's layers to be equivariant, ensuring that each layer processes symmetries in a way that is aligned with the desired group. While powerful, this approach imposes architectural constraints, which may limit the flexibility of the network and prevent leveraging large pretrained models.

\paragraph{Equivariance via canonicalization}
An alternative to weight sharing is incorporating symmetry through canonicalization \citep{kaba2022equivariance, mondal2023equivariant}, where, instead of modifying the network's architecture to handle symmetries, the input data is transformed into a canonical form. In this approach, a separate canonicalization network, which is itself equivariant, preprocesses the input, transforming it into a standard, or canonical, representation. This canonicalized input is then passed to a standard prediction network that does not need to be aware of the symmetries. If the corresponding inverse transformation is applied to the output of the prediction network, the entire model becomes equivariant; otherwise, the model remains invariant. This method has several advantages. First, it does not require altering the architecture of the prediction network, allowing for the use of large pre-trained models without modification. Second, by ensuring that the input data is in a canonical form, the prediction network only needs to learn the mapping from the canonical input to the output, without needing to learn all transformed samples. This can lead to improved performance and robustness, especially in scenarios where the prediction task does not naturally align with the symmetry group or where architectural constraints might hinder performance. Thus, in our work, we leverage canonicalization to achieve equivariance in the segmentation task. By transforming the input into a canonical form using a simple equivariant canonicalization network, we ensure that our prediction network remains unconstrained and can fully utilize its capacity for learning without the need for architectural modifications. This approach offers the benefits of symmetry-aware processing while maintaining the flexibility and power of unconstrained neural network architectures.

\section{Detailed Dataset Description}
\label{appendix:first}
\begin{figure*}[h!]
    \centering
    \includegraphics[width=0.97\linewidth]{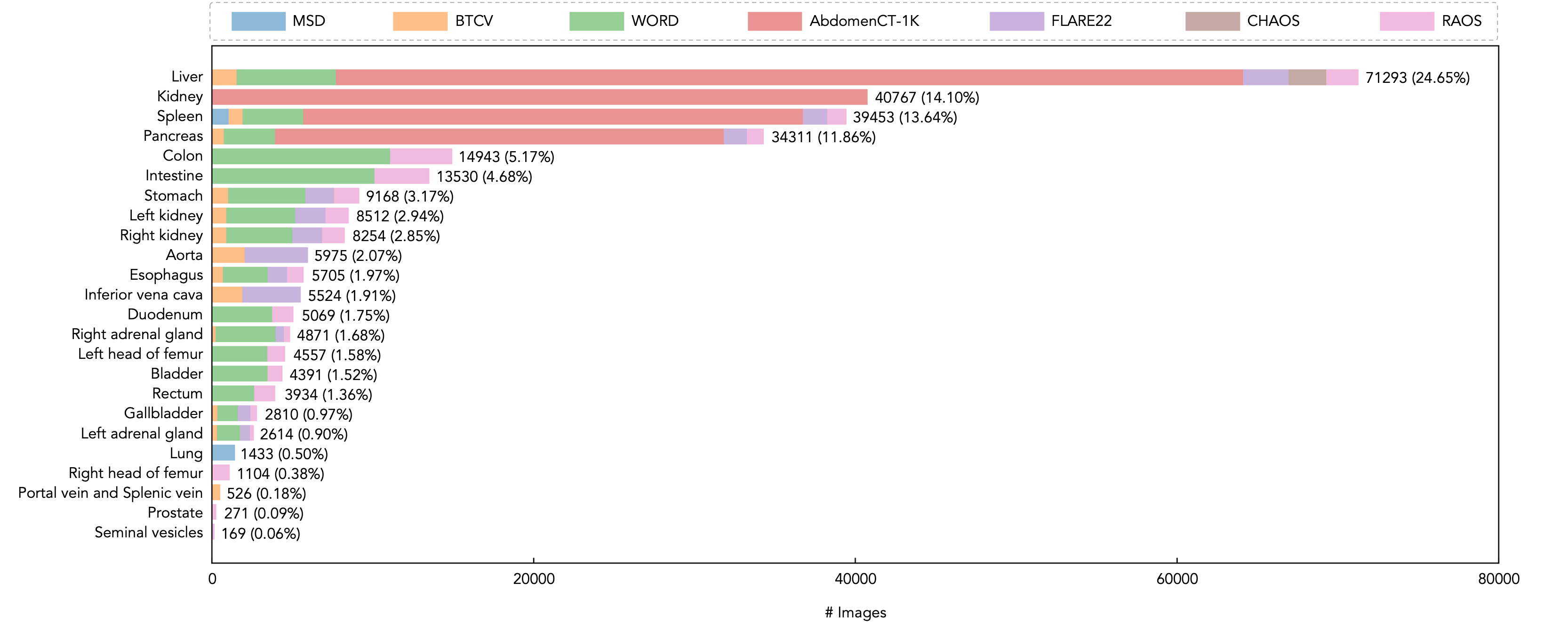}
    \caption{Distribution of labeled organs across the collected datasets. The image count for each organ and its corresponding ratio is marked in the plot.}
    \label{fig:data-statistics}
\end{figure*}

\paragraph{Image Data Collection and Preprocessing} For model development and evaluation, we collected 1,437 CT scans from 7 public datasets. A detailed summary of the datasets is provided in Table~\ref{tab:dataset}. In total, 24 organs are labled in the assembled datasets, with a strong focus on segmentation targets in the abdominal region. The organ class distribution across the datasets is shown in Fig~\ref{fig:data-statistics}. To standardize quality and reduce domain gaps, we applied a preprocessing pipeline to all datasets. Specifically, we mapped the Hounsfield unit range [-180, 240] to [0, 1], clipping values outside this range. To address dimension mismatches between datasets, masks, and images, all scans and masks were resized to $1024\times1024$. The 3D scan volumes were sliced along the axial plane to generate 2D images and corresponding masks. To ensure labeling quality, organ segments with fewer than 1,000 pixels in 3D volumes or fewer than 100 pixels in 2D slices were excluded. The finalized dataset consisted of 101,217 images, with 91,344 (90.25$\%$) used for training and validation, and 9,873 (9.75$\%$) reserved for testing.

\newcolumntype{P}[1]{>{\centering\arraybackslash}p{#1}}
\newcolumntype{L}[1]{>{\arraybackslash}p{#1}}
\begin{table*}[thb!]
\scriptsize
    \centering
    \caption{Overview of the datasets used in this study. }
    \label{tab:dataset}%
    \begin{tabular}{P{1.7cm}|P{1cm}|P{1cm}|L{8.4cm}}\toprule
         \textbf{Dataset} & \# Training scans   & \# Testing scans  &  Annotated organs\protect\footnotemark[1]  \\ \midrule
         AbdomenCT-1K	&722	&---	&Liv, Kid, Spl, Pan \\
         MSD\protect\footnotemark[2]	 &157	&---	 &Lun, Spl\\
         WORD   &100	 &20	&Liv, Spl, LKid, RKid, Sto, Gal, Eso, Pan, Duo, Col, Int, LAG, RAG, Rec, Bla, LFH, RFH\\
        FLARE22	&40	&5	& Liv, RKid, Spl, Pan, Aor, IVC, RAG, LAG, Gal, Eso, Sto, Duo, LKid \\
        CHAOS	&40	&---	&Liv\\
        BTCV &30	&---	 &Spl, RKid, LKid, Gal, Eso, Liv, Sto, Aor, IVC, PVSV, Pan, RAG, LAG\\
        RAOS\protect\footnotemark[3]	&---	&40	&
        Liv, Spl, LKid, RKid, Sto, Gal, Eso, Pan, Duo, Col, Int, LAG, RAG, Rec, Bla, LFH, RFH, Pro, SV\\
        \bottomrule
    \end{tabular}
\end{table*}

\paragraph{Test Data Creation} 

Different from existing work that solely chases for a higher segmentation accuracy, in this paper, we expect to evaluate the segment model's performance in dual tasks: The free-form text understanding ability and segmentation ability. 

\footnotetext[1]{For simplicity, the following abbreviations are used: Liv (liver), Kid (kidney), Spl (spleen), Pan (pancreas), Col (colon), Int (intestine), Sto (stomach), LKid (left kidney), RKid (right kidney), Aor (aorta), Eso (esophagus), IVC (inferior vena cava), Duo (duodenum), RAG (right adrenal gland), LHF (left head of femur), Bla (bladder), Rec (rectum), Gal (gallbladder), LAG (left adrenal gland), RHF (right head of femur), PVSV (portal vein and splenic vein), Pro (prostate), and SV (seminal vesicles).}
\footnotetext[2]{Only the  lung and spleen subsets from MSD were used.}
\footnotetext[3]{We used CancerImages (Set1) from RAOS as our out-of-domain test set. To avoid overlap, any scans in RAOS that were extended from WORD were excluded from testing.}

In order to verify the model's ability to understand the language descriptions, we construct a query dataset (test set) from two resources: 1. Real-world human queries; 2. LLM-generated synthetic queries. For the first kind of real-world queries, we have two groups of annotators, $\textbf{Domain Expert}$ and $\textbf{Non-Expert}$. Domain experts are from clinical hospitals who provide the query materials from their daily diagnosis notes, this group of people tends to use professional vocabulary, and their intention might not be explicitly expressed in a professional report, such as in the report, the doctor writes \textcolor{blue}{`Concerns in the hepatic area
that warrant a more focused
examination'}, which implicitly means the \textcolor{blue}{`liver is the area of interest under certain symptom'}. Another group of query providers is the non-expert, who are not specialized in clinical or equipped with medical specialties. We explain to this group of people that their task is to write a sentence and show the intention of segmenting the target organ/tissue in a CT scan, e.g., the liver. This aspect of real queries represents a more general and non-specialist approach to expressing the need for segmentation (such as in the student learning scenarios). Apart from real query data, we incorporate synthetic test queries to enlarge the test samples and add randomness in various expressions. The synthetic test is generated by GPT-4o following the template shown below:

\noindent\rule{0.48\textwidth}{0.4pt} 
\textbf{The Prompt Template to Generate Synthetic Queries.}

\tikz[baseline]{\draw[dashed] (0,0) -- (0.45\textwidth,0);} 

\textbf{System Description:} You are a doctor with expert knowledge of organs.

\textbf{Task Description:} Now you are making a diagnosis of a patient on the CT scan over \{\textcolor{blue}{body part}\}. You find a potential problem on \{\textcolor{blue}{organ name}\} and want to see more details in this area, please query for segmentation by free-form text. Please make sure to deliver the segment target explicitly, and you are encouraged to propose various expressions.

\textbf{Format:} \{segmentation query\}, \{explain reason\}.

\textbf{Example:} \textcolor{grey}{Given that, \{body part\} is \texttt{abdomen} and \{organ name\} is \texttt{liver}}. 

Your response should be something like: \{Please identify the liver for me for more analysis.\} \{Because elevated liver enzymes alanine aminotransferase (ALT) in the blood tests might indicate liver inflammation or damage\}.

\textbf{Output:} \{Placeholder\}

\noindent\rule{0.48\textwidth}{0.4pt} 

\begin{figure}[h!]
    \centering
    \includegraphics[width=0.99\linewidth]{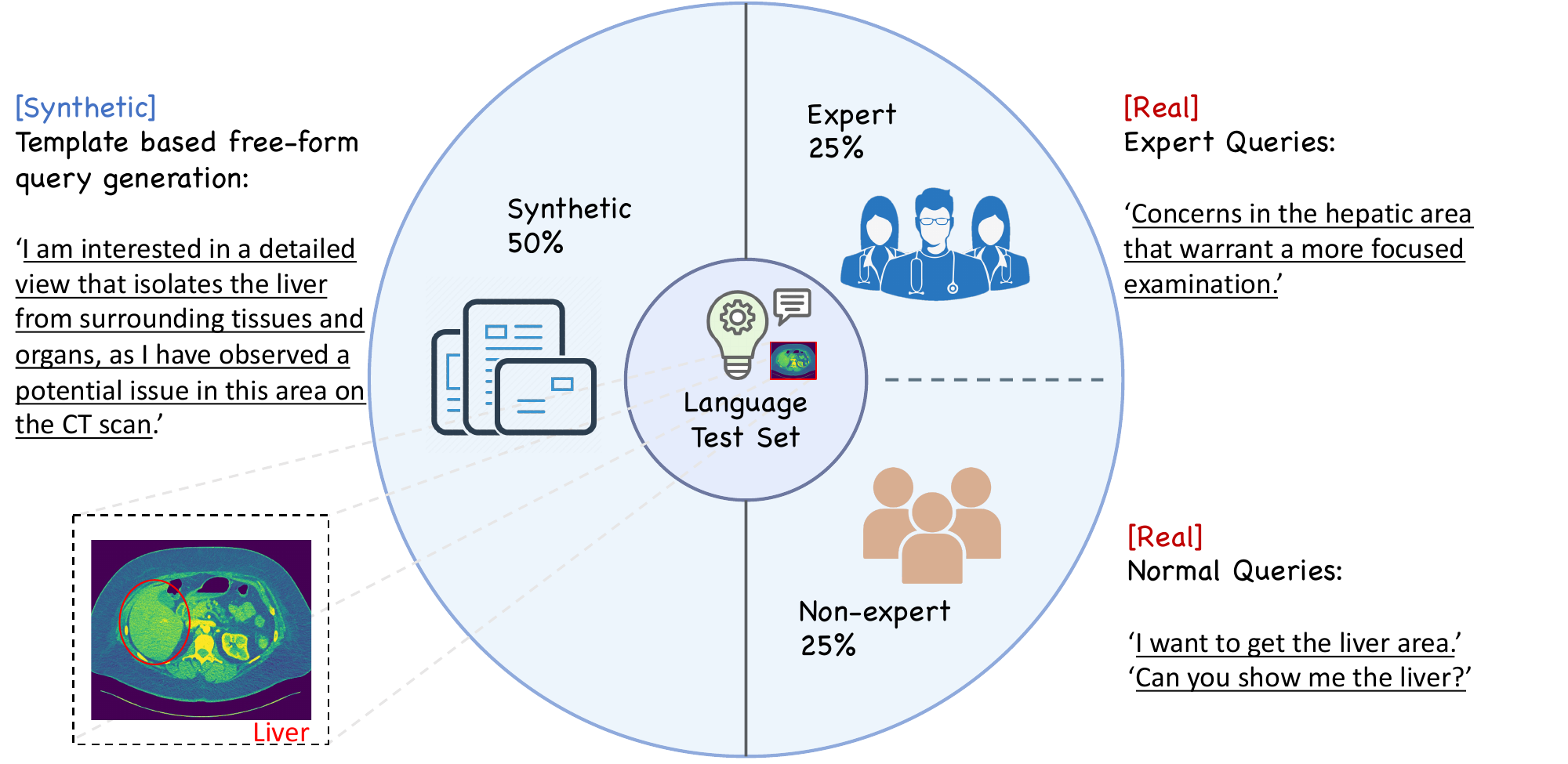}
    \caption{The Language Test Set for Verifying the Query Understanding Ability. It contains three aspects of components, real data - expert group, real data - non-expert group, and synthetic data.}
    \label{fig:testset}
\end{figure}

The overall structure of the test dataset is shown in Figure~\ref{fig:testset}. It consists of 25\% expert queries, 25\% normal queries, and half synthetic queries. In total, we have 2880 text query entries from 24 organs. Each of the queries is labeled with the correct organ name to segment. This will be used to evaluate the ability of our learned TextEncoder model to understand correct intentions based on free-form language description.
At the same time, the organ names are connected to another segmentation test set, which contains various medical images such as CT scans, MRIs, etc. And stand on the results of interest-category identification, we conduct further segmentation result analysis, including the normal segmentation precision study, and also the equivariant identified segmentation study.

\section{Training Details}\label{appendix:train}
In the training process, we provide details of the configuration files and instructions below: 

All experiments were conducted on an AWS ml.p3dn.24xlarge instance equipped with 8 V100 GPUs, each with 32 GB of memory. We used a batch size of 16 and applied the CosineAnnealingLR learning rate scheduler, initializing the learning rate for all modules at 0.0001. The AdamW optimizer was employed for training. A small $D_8$ equivariant canonicalization network was used, consisting of 3 layers, a hidden dimension of 8, and a kernel size of 9. To maintain consistency across the input and output formats, all scans from different datasets were resized to 1024$\times$1024 and both predicted and ground truth masks were resized to 256$\times$256 for fair comparison. For images with a single channel, the channel was duplicated to 3. All models’ performance on the test sets is reported using both the Dice coefficient \citep{taha2015metrics} and normalized surface distance following the existing work \citep{heimann2009statistical}.

\section{Attention Map}
In this section, we demonstrate the attention map to visualize the correlation between text embedding and image embedding.

These attention maps based on autonomy-agnostic prompts not only address the first reviewer's query about plotting the relationship between text and images but also counter the second reviewer's concern that 'our method seems capable of inferring which organ is desired given a text prompt and images.

For example, consider the attention map generated with the prompt `segment the part located at the topmost portion' it does not highlight just one organ. Instead, all organs at the top are highlighted. This demonstrates that our model is not merely overfitting data to infer a specific organ; rather, it has a deep understanding of the text and its relationship to the medical image.

\begin{figure*}[h!]
    \centering
    \includegraphics[width=0.99\linewidth]{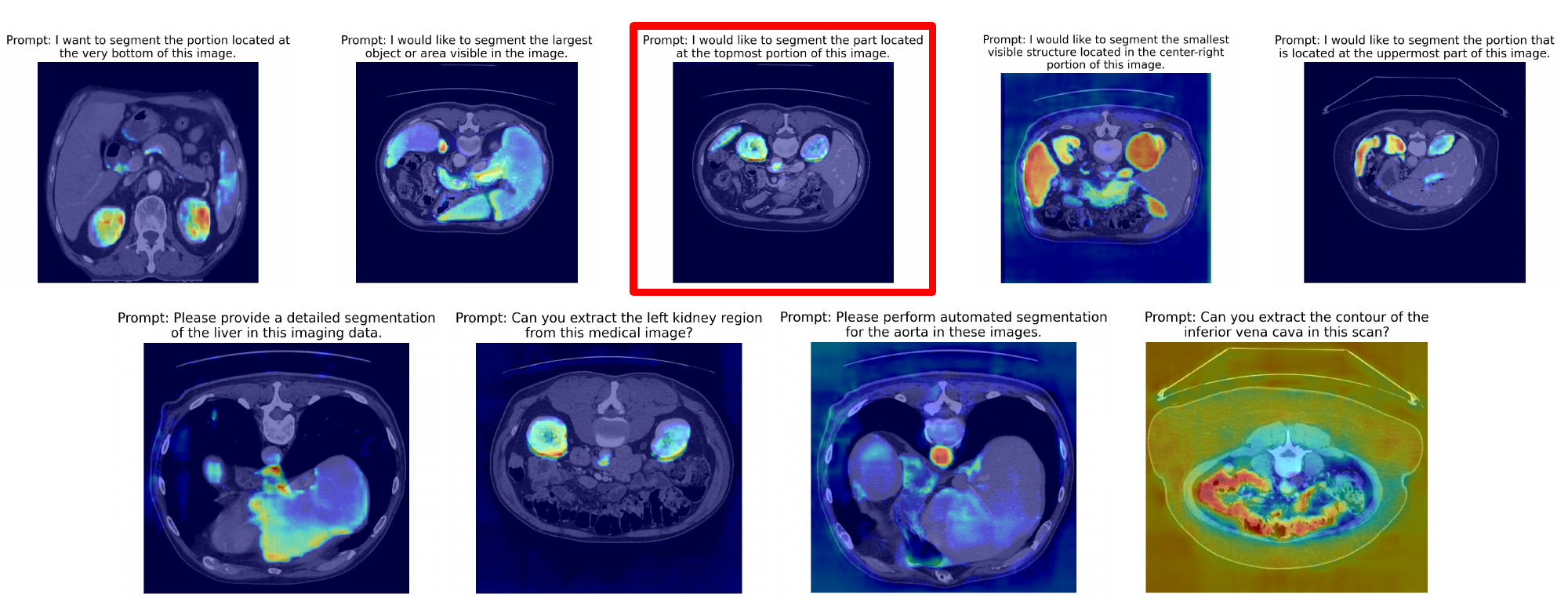}
    \caption{The examples of attention maps in \ours segmentation tasks. The attention maps are computed based on the scaled product of text embedding from the text encoder and image embedding from the image encoder.}
    \label{fig:enter-label}
\end{figure*}

\section{More Use Case Demonstrations}
\begin{figure*}
    \centering
    \includegraphics[width=0.90\linewidth]{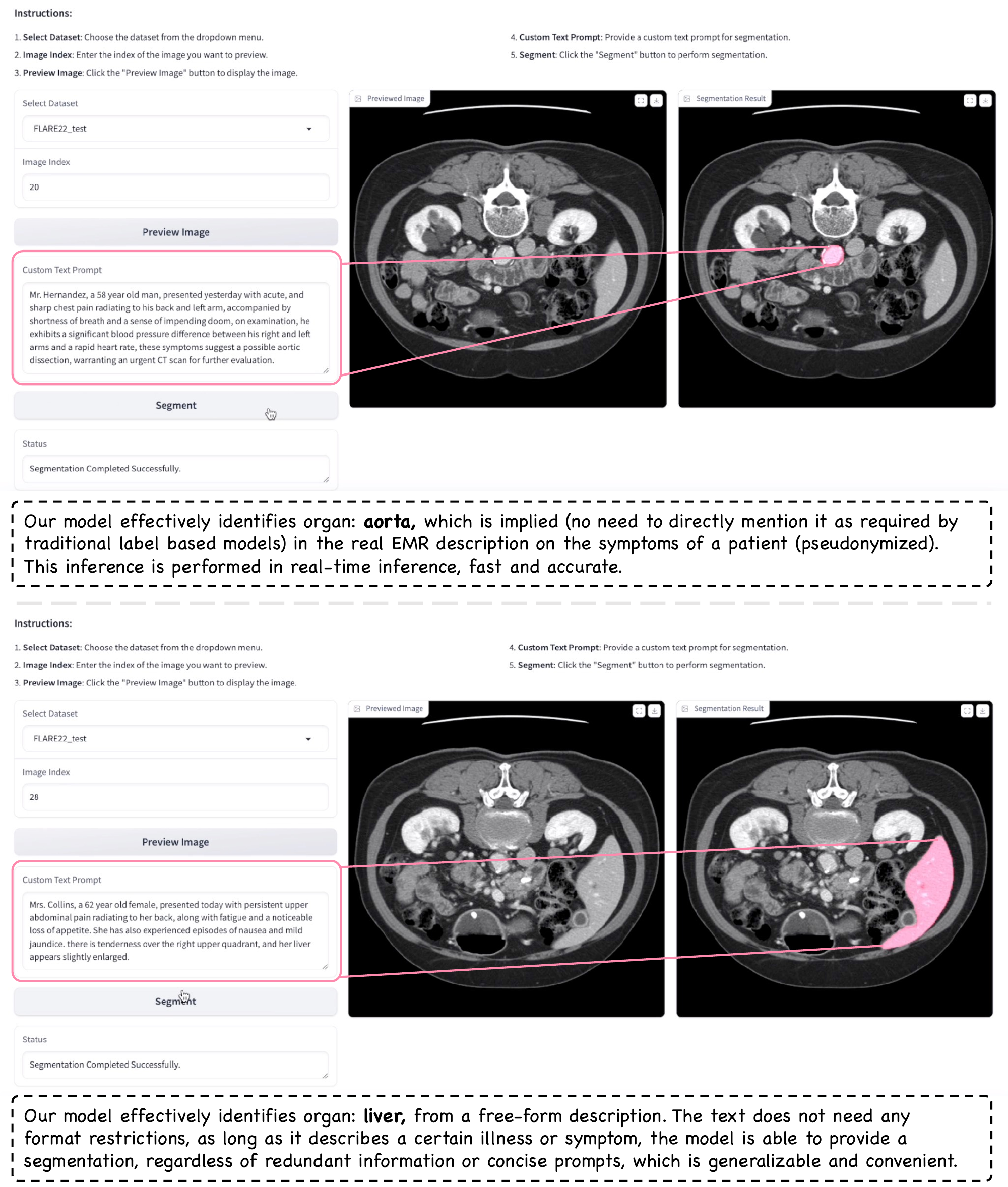}
    \caption{The implicit Anatomy-Informed demonstration on a deployed version of the \ours model, it provides real-time inference ability and can be robust to any format type of the prompts, either lengthy or redundant, it can still perform effectively to identify the most cared organ. In this image, two examples of real EMR record data are provided (personal information such as name and age is pseudonymized.) We could observe that, even though sometimes the actual label is not explicitly described as in the upper half of this image, the model can perform segmentation accurately because it has seen the semantic similar corpus in the training period and aligned these symptom-related texts with the correct segmentation area.}
    \label{fig:demo1}
\end{figure*}

\begin{figure*}
    \centering
    \includegraphics[width=0.90\linewidth]{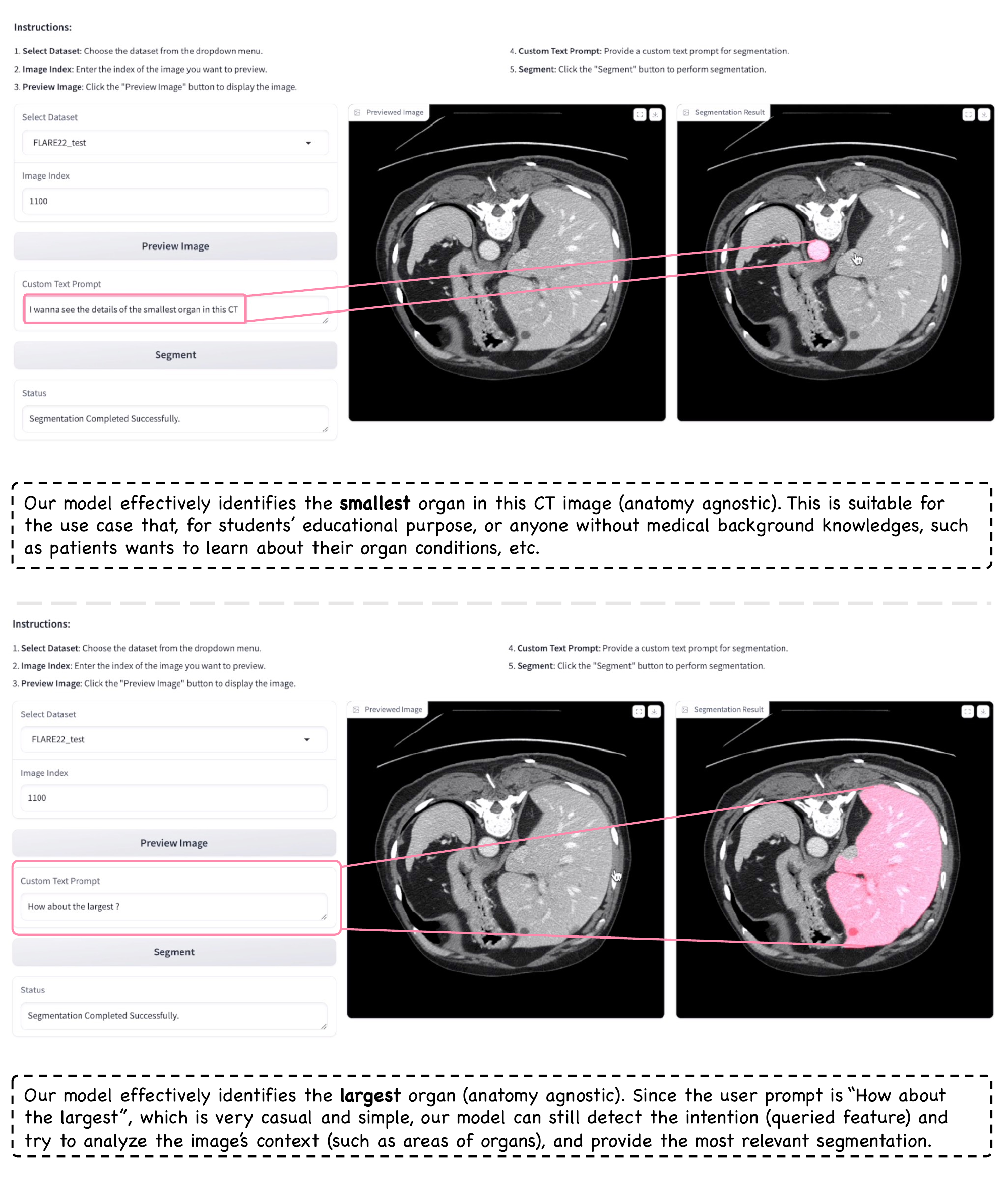}
    \caption{The Anatomy-Agnostic demonstration on a deployed version of the \ours model, this feature is specially designed for a larger group of users who care about the medical image scannings, but lack the professional background knowledge, such as students, and patients. This supports the segmentation with the size relevant or positional relevant prompts, the above figures show that \ours is able to successfully segment those largest, sand mallest organs in the provided scans. }
    \label{fig:demo2}
\end{figure*}

\begin{figure*}
    \centering
    \includegraphics[width=0.95\linewidth]{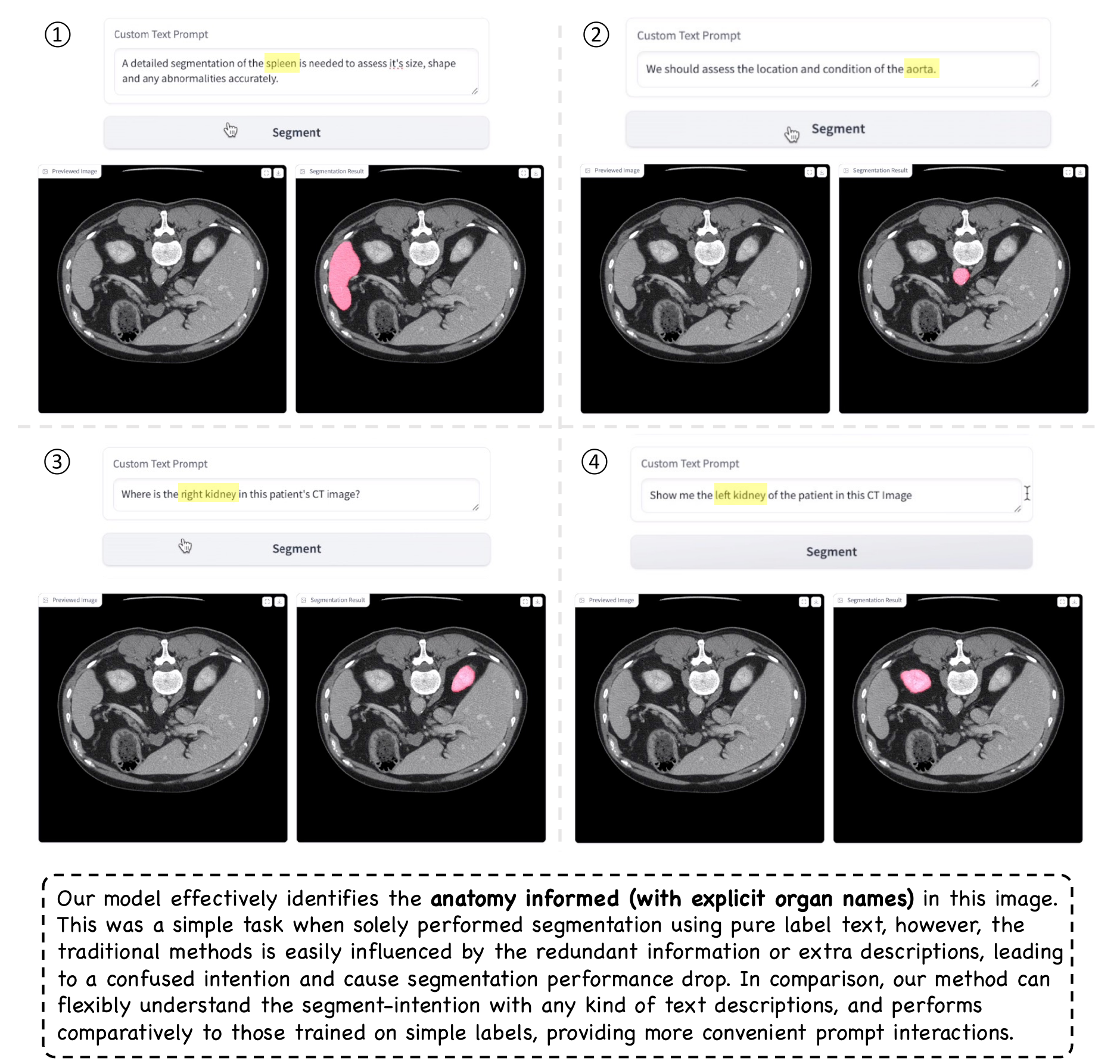}
    \caption{The explicit Anatomy-Informed demonstration on a deployed version of the \ours model, in this figure, we showcase four examples of segmentation results by organ names relevant prompts. In the prompt content, it mentions the organ names, so as to instruct the segment action. As in the \ding{172} and \ding{173}, the model segment \texttt{spleen} and \texttt{aorta} successfully. And as shown in the \ding{174} and \ding{175}, the model is also able to distinguish the \texttt{right kidney} and \texttt{left kidney} regardless of the angle and position of the scan is taken, enforced by the canonicalization module as introduced in the Section~\ref{sec:cano}.}
    \label{fig:demo3}
\end{figure*}

In this section, we provide more examples of segmentation results, ranging from the Anatomy-Informed to Anatomy-Agnostic ones. In Anatomy-Informed segmentation, we conduct two versions of illustration, first, we show the simplest (organ name is explicitly described in the prompt) segmentation, as shown in Fig.~\ref{fig:demo3}, and to add on more complexity and showcase how \ours is beneficial for real-world clinical use cases, we take the diagnosis data from pseudonymized real EMR (electronic medical record) data in Fig.~\ref{fig:demo1}, \ours is able to detect the organ from the long and redundant descriptive symptoms and provide accurate segmentation, this is especially useful for providing diagnosis assistance based on doctors' notes. And last, we show size late Anatomy-Agnostic segmentations as in Fig.~\ref{fig:demo2}.

\begin{figure*}
    \centering
    \includegraphics[width=0.90\linewidth]{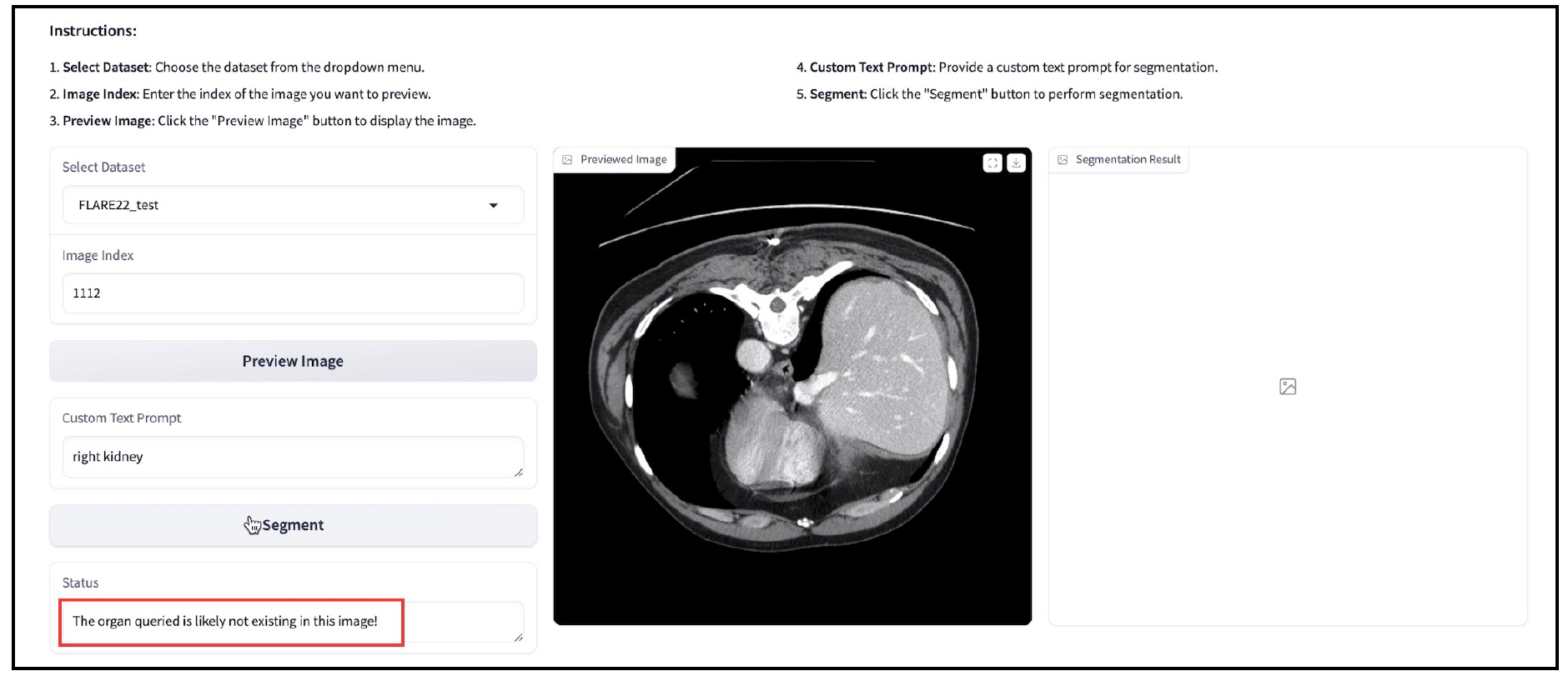}
    \caption{This image provides an extra analysis on the corner case detection. As shown in this image, there exist situations in blurry, low-quality metadata leading to the invisibility of an organ, or that an organ does not exist. \ours is able to detect such cases and provide feedback that \textit{The organ queried is likely not existing in this image!}. This is realized by a filter layer of function upon the predicted $\text{Probability}$ for an organ area, the threshold is set to $\alpha = 0.5$. One can easily tune the parameter based on the actual required confidence of the \ours.}
    \label{fig:demo_corner}
\end{figure*}

\end{document}